\newlength{\intwidth}

\documentclass{jfm}
\usepackage{amssymb}

\usepackage{array}
\usepackage{amsmath}
\usepackage{latexsym}
\usepackage{bezier}
\usepackage{psfrag,graphicx}
\usepackage{bm}
\usepackage{float}
\usepackage{multirow}
\usepackage{mathrsfs}
\usepackage{color}
\newcommand{\Real}{\mathbb{R}}

\newtheorem{thm}{Theorem}

\begin{document}

\title[Eigenvalue bounds for magneto-shear flows]{Eigenvalue bounds for compressible stratified magneto-shear flows varying in two transverse directions}
\author{Kengo Deguchi}
\affiliation{
School of Mathematics, Monash University, VIC 3800, Australia
}

\maketitle

\begin{abstract}
Three eigenvalue bounds are derived for the instability of ideal compressible stratified magnetohydrodynamic shear flows in which the base velocity, density, and magnetic field vary in two directions. The first bound can be obtained by combining the Howard semicircle theorem with the energy principle of the Lagrangian displacement. 
%
%
Remarkably, no special conditions are needed to use this bound, and for some cases, we can establish the stability of the flow. 
The second and third bounds come out from a generalisation of the Miles-Howard theory and have some similarity to the semi-ellipse theorem by Kochar \& Jain (\textit{J. Fluid Mech.}, vol. 91, 1979, 489) and the bound found by Cally (\textit{Astrophys. Fluid Dyn.}, vol. 31,1983, 43), respectively.
An important byproduct of this investigation is that the Miles-Howard stability condition holds only when there is no applied magnetic field and, in addition, the directions of the shear and the stratification are aligned \textcolor{black}{everywhere}. 
%
\end{abstract}

\section{Introduction}

This paper aims to find \textit{a priori} complex growth rate bounds of ideal wave-like perturbations on top of compressible, stratified, magnetised, and sheared equilibrium states varying in two directions perpendicular to the direction of the wave propagation. 

Growth rate bounds and stability criteria of ideal shear flows have long been one of the central issues in theoretical fluid mechanics. Rayleigh's inflection point theorem (Rayleigh 1880), Howard's semicircle theorem (Howard 1961), and the Miles-Howard stability condition (Miles 1961) are particularly well-known across the fluid dynamics, geophysics, and astrophysics communities, and their various extensions have been sought. 
One obvious way of extending the theories is to make the base state variation two-directional as in many practical problems. However, far less progress has been made for the theoretical understanding of such generalised stability problems compared with the classical cases, in particular when there are multiple physical effects such as compressibility, stratification, and magnetohydrodynamic effects. The major difference between the planar and non-planar cases is that for the former classical cases the motion of the perturbation might be restricted on a two-dimensional plane. In the geophysics community, the stability theory has been advanced using the fact that multiple Casimir invariants are available for two-dimensional fluid motion when it is viewed in the Hamiltonian mechanics framework (see Arnold (1978), Dowling (1995), for example). While for the generalised problem the perturbation is inherently three-dimensional, and so fewer mathematical tools are available. 
Whether the classical planar results can be carried over to non-planar cases is in general a non-trivial question.

The non-planar magneto-hydrodynamic stability problems may have some relevance to cutting-edge industrial applications such as fusion reactors or magneto-hydrodynamic electric generators.
For example, the magnetohydrodynamic flows through ducts of complicated cross-sections commonly appear in some fusion blanket designs. 
%
The Hunt flow in a rectangular duct may be one of the simplest model for this problem, where the inviscid instability indeed plays important roles (Priede et al. 2010; Qi et al. 2017). 
Even in such a simple case, the stability computation is not easy, especially in the high Reynolds number range. 
Moreover, when the duct has sharp corners the singularities there affect the numerical eigenvalues badly. If fins of complex shape are attached to enhance heat transfer, the computation becomes practically impossible.

%
The flow fields to be treated in this paper may also have important implications for solar physics. 
For example, just underneath the Sun's surface,
there is a shear layer called \textit{tachocline} (see Charbonneau et al. 1998, for example)
and the stability of it has been a subject of many astrophysics studies, wherein the planar version of our base flow was commonly used. 
%
In the early years of the stability analyses in solar physics, the effect of shear was omitted because in that case the stability can be analysed by simply examining the property of the Lagrangian potential energy (Bernstein et al. (1958), Newcomb (1961), Parker (1966)). Frieman \& Rotenberg (1960) introduced the effect of shear to the energy principle theory, and based on the extended principle, Adam (1978a) and Tobias \& Hughes (2004) subsequently developed some stability conditions; we will add further comments on those studies later.

Here, we summarise the previous (mostly planar) stability results that are relevant to our results. 
The semicircle theorem by Adam (1978b) is particularly relevant to the present study because it is shown for the two-dimensional version of the magneto-atmospheric flow configuration to be considered in this paper. In the first half of this paper we will extend the application range of his result significantly and, in addition, improve the eigenvalue bound itself. 
Hughes \& Tobias (2001) found that there are two semicircles possible, and union of them will give a net bound. Motivated by this result, we shall show that actually infinitely many semicircle bounds can be defined, and the union of them constitutes the eigenvalue bound better than the usual semicircle one. Another improvement shown in Hughes \& Tobias (2001) is that the semicircle radius can be contracted when the external magnetic field presents. The same result was shown in Cally (2000), who found that the contraction discovered in Howard \& Gupta (1962) and Gupta (1992) for a uniform magnetic field case can be carried over to inhomogeneous magnetic fields. The stabilisation effect of the magnetic field may also be seen for compressible flows, in view of the stability condition found by Cally (2000). This is in fact true as we shall show in this paper, although there is a condition for the stabilisation to occur.

In the latter half of this paper we shall derive two other types of bounds using 
the Miles-Howard theory. The theory differs fundamentally from the semicircle type theories in that the stability condition deduced by it depends on the velocity shear, rather than the velocity range. Our second bound is similar to the semi-ellipse bound by
Kochar \& Jain (1979), who found that the inequality derived from the Miles-Howard theory can be used to improve the semicircle eigenvalue bound. 
Fung (1986) attempted to extend the semi-ellipse theorem to 
 flows subjected to a general conservative force field varying in two directions.
\textcolor{black}{However, Fung (1986) implicitly assumed a condition that can only be satisfied for quite specific flows.}
Our bound is found by considering the correct extension of the semi-ellipse theorem, even including the compressibility and magneto-hydrodynamic effects, that were not considered in Kochar \& Jain (1979) and Fung (1986).
The third bound to be found in this paper is the non-planar version of the bound found by Cally (1983) which also depends on the velocity shear. 
As Cally's derivation cannot be extended to the generalised flows, we instead use the Miles-Howard theory in the proof.


Among all the above previous stability results, the incompressible result by Gupta (1992) is the only work where a magnetised problem was treated in the base shear varying in two directions. 
For non-magnetised flows, to the best of the author's knowledge, Eckart (1963) and Hocking (1964) were the first to consider such an extension, with the former showing that Howard's semicircle theorem holds for some cases and the latter leading to limited results for special flow fields, respectively. 
The work by Eckart (1963) seems to be often cited in the context of adding compressibility to the semicircle theorem, and the aspect of dealing with a flow that varies in two directions is less well known. Perhaps, for this reason, many authors subsequently reported semicircle theorems that are covered by Eckart's result (e.g. Blumen (1975) for convectively stably stratified Boussinesq flows, Dandapat \& Gupta (1977) for compressible, unstratified flows, Li (2011) and Waleffe (2019) for incompressible, unstratified flows).

\textcolor{black}{
Numerical stability analyses of inviscid flows varying in two directions are in general very challenging, unlike the planar classical planar versions for which we can use the shooting method.
For the simplest case where the flow is incompressible, non-magnetised and unstratified, Hall \& Horseman (1991), Yu \& Liu (1991), Li \& Malik (1995) and Andersson et al. (2001) 
successfully computed the stability of the streak (which of course varies in two directions, because it refers heterogeneity in the velocity field created by weak vortices elongated in the flow direction).
However, for flows with more complex physical effects, such as those considered in this paper, 
no numerical results has been reported.
%
The significance of this paper is that we have developed a quick and simple method to study such challenging eigenvalue problems. The bounds to be derived can be easily calculated for a quite wide range of flow configurations to see the overall character of the eigenvalues. 
A particularly important practical engineering application would be the stability problem of the aforementioned magnetohydrodynamic flows through ducts with complex cross-section. 
}

\textcolor{black}{
Although ideal stability analyses might seem somewhat classical, the importance of it in  turbulence is increasingly recognised in the fluid dynamics community. 
There are much evidence that the instability waves on top of the streak indeed 
play crucial roles in fully developed near-wall turbulent flows
(Hamilton et al. 1995; Jimenez \& Pinelli 1999; McKeon \& Shama 2010; Thomas et al. 2014; Beaume et al. 2015; Alizard 2015). 
If we accept this view, the complex eigenvalue bounds should have much implication for the properties of turbulent bursts.
Furthermore, when the amplitude of the instability waves of the streak reaches a certain magnitude, they eventually drive the streak field via Reynolds stress (see Benney 1984; Hall \& Smith 1991; Waleffe 1997). The resultant nonlinear loop between the streak and the waves constitutes the basis of the self-sustainment mechanism of coherent structures in near wall turbulence and has recently attracted much attention.
Interestingly, at high Reynolds numbers the mechanism by which the instability wave of the streak is generated can be well described by the inviscid theory. Indeed, the fact that the generalised ideal stability problem sits at the centre of the sustainment process inspired the recent two re-discoveries of the semicircle theorem (Li 2011, Waleffe 2019). }

The paper is organised as follows. Section 2 formulates the problem based on the ideal compressible magnetohydrodynamic equations under the influence of the general conservative force field. We will show that the linear stability problem can be reduced to three equations for the Lagrangian displacement vector, or alternatively, a single equation for the total pressure perturbation. 
In section 3, we will first describe the derivation of the semicircle bound based on the energy principle theory, and then see how the bound can be improved by more effectively using the property of the potential energy. 
The primal focus of section 4 is to clarify when the Miles-Howard theory can be used for the generalised non-planar flows.
The results obtained in the generalised theory will then be used to develop a semi-ellipse type bound and a Cally (1983) type bound, both of which are generalised for the non-planar flows.
The three bounds are compared with numerical eigenvalues in section 5.
Finally, in section 6, we will draw some conclusions.


\section{Formulation of the problem}

\subsection{Derivation of the stability problem}
Consider the ideal compressible magneto-hydrodynamic equations in the cartesian coordinates $(x,y,z)$. 
\begin{subequations}\label{original}
\begin{eqnarray}
\rho(\partial_{t}+\mathbf{v}\cdot \nabla)\mathbf{v}
=-\nabla q+\frac{1}{\mu_0}(\mathbf{b}\cdot \nabla)\mathbf{b}+\rho \nabla G,\\
\partial_{t}\mathbf{b} 
= \nabla \times (\mathbf{v}\times \mathbf{b}),\\
\partial_{t} \rho+ \nabla \cdot (\rho \mathbf{v})=0,\\
(\partial_{t}+\mathbf{v}\cdot \nabla)(\rho^{-\gamma} p)=0.
\end{eqnarray}
\end{subequations}
The first set of equations are the conservation of the momentum, the second set of equations are the induction equations, the third equation is the conservation of the mass, and the fourth equation is the adiabatic energy equation.
In addition, we assume that Gauss's law $\nabla \cdot \mathbf{b}=0$ holds.
Throughout the paper, we denote the velocity vector as $\mathbf{v}=(v^x,v^y,v^z)$, the magnetic field vector as $\mathbf{b}=(b^x,b^y,b^z)$, the density as $\rho$, the kinematic pressure as $p$, and the total (i.e. kinematic and magnetic) pressure as $q=p+|\mathbf{b}|^2/2\mu_0$.
The momentum equations are subjected to a conservative field with potential $G(y,z)$, as in Fung (1986).
The adiabatic exponent $\gamma$ and the vacuum permeability $\mu_0$ are constants.


Our interest is the stability of a quasi-equilibrium base state depending on $y,z$. 
The base velocity $\mathbf{v}=(\overline{u}(y,z),0,0)$ and magnetic field $\mathbf{b}=(\overline{B}(y,z),0,0)$ are assumed to be unidirectional. 
Those fields may be driven by some external forcing applied on the streamwise component of the momentum and induction equations, or may be developing in slower time scale than that of the ideal instability.
From $y,z$ components of the momentum equations, it is easy to see that the base total pressure $\overline{q}(y,z)$ and the base density $\overline{\rho}(y,z)$ must satisfy the magneto-static conditions 
\begin{eqnarray}
\overline{q}_y-\overline{\rho}G_y=\overline{q}_z-\overline{\rho}G_z=0,\label{magstat}
\end{eqnarray}
which imply
\begin{eqnarray}
\frac{G_y}{\overline{\rho}_y}=\frac{G_z}{\overline{\rho}_z}.\label{FUNGcond}
\end{eqnarray}
Here and hereafter the subscripts $y$ and $z$ represent corresponding partial differentiation.

The stability of the base flow can be found by adding an infinitesimally small normal mode perturbation:
\begin{subequations}\label{expand}
\begin{eqnarray}
\mathbf{v}=
\left [
\begin{array}{c}
\overline{u}(y,z)\\0\\0
\end{array}
\right ]
+
\left [
\begin{array}{c}
\widetilde{v}^x(y,z)\\ \widetilde{v}^y(y,z)\\ \widetilde{v}^z(y,z)
\end{array}
\right ]e^{ik (x-ct)}+\text{c.c.},\\
\mathbf{b}=
\left [
\begin{array}{c}
\overline{B}(y,z)\\0\\0
\end{array}
\right ]
+
\left [
\begin{array}{c}
\widetilde{b}^x(y,z)\\ \widetilde{b}^y(y,z)\\ \widetilde{b}^z(y,z)
\end{array}
\right ]e^{ik(x-ct)}+\text{c.c.},
\end{eqnarray}
\begin{eqnarray}
\rho=\overline{\rho}(y,z)+\widetilde{\rho}(y,z)e^{ik (x-ct)}+\text{c.c.},\\
p=\overline{p}(y,z)+\widetilde{p}(y,z)e^{ik (x-ct)}+\text{c.c.},\\
q=\overline{q}(y,z)+\widetilde{q}(y,z)e^{ik (x-ct)}+\text{c.c.},
\end{eqnarray}
\end{subequations}
where the Fourier transformed perturbation quantities $\widetilde{v}^x,\widetilde{v}^y,\widetilde{v}^z,\widetilde{b}^x,\widetilde{b}^y,\widetilde{b}^z,\widetilde{\rho},\widetilde{p},\widetilde{q}$ are complex functions and c.c. stands for complex conjugate.
Here $k>0$ is the streamwise wavenumber and $c=c_r+ic_i$ is the complex wave speed.
We remark here that 
from the definition of the total pressure, the base and perturbation kinetic pressures should satisfy $\overline{p}=\overline{q}-\overline{B}^2/2\mu_0$ and $\widetilde{p}=\widetilde{q}-\overline{B}\widetilde{b}^x/\mu_0$, respectively.

Substituting (\ref{expand}) to (\ref{original}) and neglecting all the nonlinear terms, we find
\begin{subequations}\label{full_linear}
\begin{eqnarray}
\overline{\rho}
\left \{
Uik
\left[ \begin{array}{c} \widetilde{v}^x\\ \widetilde{v}^y\\ \widetilde{v}^z \end{array} \right] 
+
\left[ \begin{array}{c} \widetilde{v}^yU_y+\widetilde{v}^zU_z\\ 0\\ 0 \end{array} \right]
\right \} \hspace{60mm}\nonumber \\
-
\frac{1}{\mu_0}\left \{
\overline{B}ik
\left[ \begin{array}{c} \widetilde{b}^x\\ \widetilde{b}^y\\ \widetilde{b}^z \end{array} \right] 
+
\left[ \begin{array}{c} \widetilde{b}^y\overline{B}_y+\widetilde{b}^z\overline{B}_z\\ 0\\ 0 \end{array} \right]
\right \}
+\left[ \begin{array}{c} ik \widetilde{q} \\ \widetilde{q}_y-G_y\widetilde{\rho}  \\ \widetilde{q}_z-G_z\widetilde{\rho} \end{array} \right]=0,~~~~\label{comp_mo}
\end{eqnarray}
\begin{eqnarray}
\left \{
Uik
\left[ \begin{array}{c} \widetilde{b}^x\\ \widetilde{b}^y\\ \widetilde{b}^z \end{array} \right] 
+
\left[ \begin{array}{c} \widetilde{v}^y\overline{B}_y+\widetilde{v}^z\overline{B}_z\\ 0\\ 0 \end{array} \right]
\right \} \hspace{70mm} \nonumber \\
-
\left \{
\overline{B}ik
\left[ \begin{array}{c} \widetilde{v}^x\\ \widetilde{v}^y\\ \widetilde{v}^z \end{array} \right] 
+
\left[ \begin{array}{c} \widetilde{b}^yU_y+\widetilde{b}^zU_z\\ 0\\ 0 \end{array} \right]
\right \}
+
\left[ \begin{array}{c} \overline{B}(ik \widetilde{v}^x+\widetilde{v}_y^y+\widetilde{v}_z^z)\\ 0\\ 0 \end{array} \right] 
=0,
\label{comp_ind}~~~~ 
\end{eqnarray}
\begin{eqnarray}
ik \widetilde{b}^x+\widetilde{b}_y^y+\widetilde{b}_z^z=0, ~~~~~~~~~~\label{comp_div}\\
U ik \widetilde{\rho}+(\widetilde{v}^y\overline{\rho}_y+\widetilde{v}^z\overline{\rho}_z)+\overline{\rho}(ik \widetilde{v}^x+\widetilde{v}_y^y+\widetilde{v}_z^z)=0,~~~~~\label{comp_mass} \\
U ik  (s^2 \widetilde{\rho}+\frac{\overline{B}\widetilde{b}^x}{\mu_0}-\widetilde{q})+\widetilde{v}^y(s^2 \overline{\rho}_y+\frac{\overline{B}\, \overline{B}_y}{\mu_0}-G_y\overline{\rho})+\widetilde{v}^z(s^2 \overline{\rho}_z+\frac{\overline{B}\, \overline{B}_z}{\mu_0}-G_z\overline{\rho})=0.~~~~~\label{comp_state}
\end{eqnarray}
\end{subequations}
Here the local sound wave speed $s$ and the shifted base velocity $U$ have been defined as
\begin{eqnarray}
s(y,z)\equiv \sqrt{\frac{\gamma\overline{p}}{\overline{\rho}}}, \qquad U(y,z)\equiv \overline{u}(y,z)-c,
\end{eqnarray}
respectively.


\subsection{The stability equations in terms of Lagrangian displacement}
In order to simplify the linearised equations, following Frieman \& Rotenberg (1960), we introduce the Lagrangian displacement $(\xi,\eta,\zeta)$ such that
\begin{eqnarray}
\widetilde{v}^x=ik U\xi-U_y \eta -U_z \zeta ,\qquad \widetilde{v}^y=ik U \eta,\qquad \widetilde{v}^z=ik U \zeta.
\end{eqnarray}
Upon using (\ref{comp_ind}), (\ref{comp_mass}), and (\ref{comp_state}), it is easy to see that the other wave variables are also written in terms of the displacement as
\begin{subequations}
\begin{eqnarray}
\widetilde{b}^x=-(\overline{B}_y \eta +\overline{B}_z \zeta) -\overline{B}(\eta_y+\zeta_z),\qquad \widetilde{b}^y=ik \overline{B} \eta,\qquad \widetilde{b}^z=ik \overline{B}\zeta,~~~~\\
\widetilde{\rho}=-(\overline{\rho}_y \eta +\overline{\rho}_z \zeta) -\overline{\rho}(i\alpha \xi+\eta_y+\zeta_z),\label{1rho}\\
\widetilde{q}=-s^2\overline{\rho}(ik \xi+\eta_y+\zeta_z)-\frac{\overline{B}^2}{\mu_0}(\eta_y+\zeta_z)-\overline{\rho}(G_y\eta+G_z\zeta),~~\label{2q}
\end{eqnarray}
\end{subequations}
while the momentum equations (\ref{comp_mo}) can be transformed into
\textcolor{black}{
\begin{subequations}\label{momxi}
\begin{eqnarray}
%
%
%
%
k ^2\overline{\rho}U^2 \xi=ik \widetilde{q}+ik \overline{\rho}a^2(\eta_y+\zeta_z),\label{2xi}\\
k^2\overline{\rho}U^2\eta=\widetilde{q}_y+k^2\overline{\rho}a^2\eta -G_y\widetilde{\rho},\label{1eta}\\
k^2\overline{\rho}U^2\zeta=\widetilde{q}_z+k^2\overline{\rho}a^2\zeta-G_z\widetilde{\rho}.\label{1zeta}
\end{eqnarray}
\end{subequations}
}
Here $a(y,z)$ is the local Alfv\'en wave speed
\begin{eqnarray}
a\equiv \frac{\overline{B}}{\sqrt{\mu_0\overline{\rho}}}.\label{defAlf}
\end{eqnarray}
Eliminating $\widetilde{\rho}$ and $\widetilde{q}$ from (\ref{momxi}) using (\ref{1rho}), (\ref{2q}), we have the three equations for the displacement which is often written in the form (see Bernstein et al. (1958), Frieman \& Rotenberg (1960))
\textcolor{black}{
\begin{eqnarray}\label{dispstab}
\overline{\rho}k^2c^2
\left [
\begin{array}{c}
\xi\\ \eta\\ \zeta
\end{array}
\right ]-2\overline{\rho}k^2c\overline{u}
\left [
\begin{array}{c}
\xi\\ \eta\\ \zeta
\end{array}
\right ]+\mathbf{F}(\xi,\eta,\zeta)=\mathbf{0}.\label{xiforce}
\end{eqnarray}
}
The self-adjoint operator $\mathbf{F}$ acting on the displacement is called a force operator in the theoretical magnetohydrodynamics community. 
Its explicit form is rather complicated, and involves the Alfv\'en wave speed $a$, the sound wave speed $s$, and  the local buoyancy (Brunt-V\"ais\"al\"a) frequencies
\begin{eqnarray}
N_1^2\equiv \frac{G_y\overline{\rho}_y}{\overline{\rho}}-\frac{G_y^2}{s^2},~~~
N_2^2\equiv \frac{G_z\overline{\rho}_z}{\overline{\rho}}-\frac{G_z^2}{s^2},~~~
N_{12}^2\equiv \frac{G_z\overline{\rho}_y}{\overline{\rho}}-\frac{G_yG_z}{s^2},\label{N12}
\end{eqnarray}
which are similar to those used in Fung (1986), but here the compressible effect is included.

The linear stability problem (\ref{full_linear}), equivalent to the displacement equations (\ref{xiforce}), are sought in a domain $\Omega \subset \Real^2$ to seek the eigenvalue $c$. The boundary $\partial \Omega$ are assumed to be made of periodic boundaries and/or impermeable boundaries at which the derivative of $\widetilde{q}$ normal to $\partial \Omega$, and the normal component of the displacement vector, must vanish. 
Essentially, those assumptions are needed to eliminate the boundary terms in the various integration by parts to be seen in the subsequent sections. The eigenvalue bounds in this paper are also valid even when a free surface is introduced, i.e. $\widetilde{q}=0$ at the boundary.


An alternative common way to reduce ideal governing equations is to combine them into a single equation for the total pressure (Hocking (1968); Goldstein (1976), Benney (1984); Henningson (1987); Hall \& Horseman (1991), in particular it is used 
by Li (2011) to re-derive the semicircle theorem).
This is still possible for our system (\ref{full_linear}), although the system we are dealing with contains many physical effects (see Appendix A).
However, the equation for $\widetilde{q}$ obtained in this manner is very complicated, 
and is not so useful in the theoretical analysis of our problem.

\subsection{The model flow}

\begin{figure}
\centering
\includegraphics[scale=0.45]{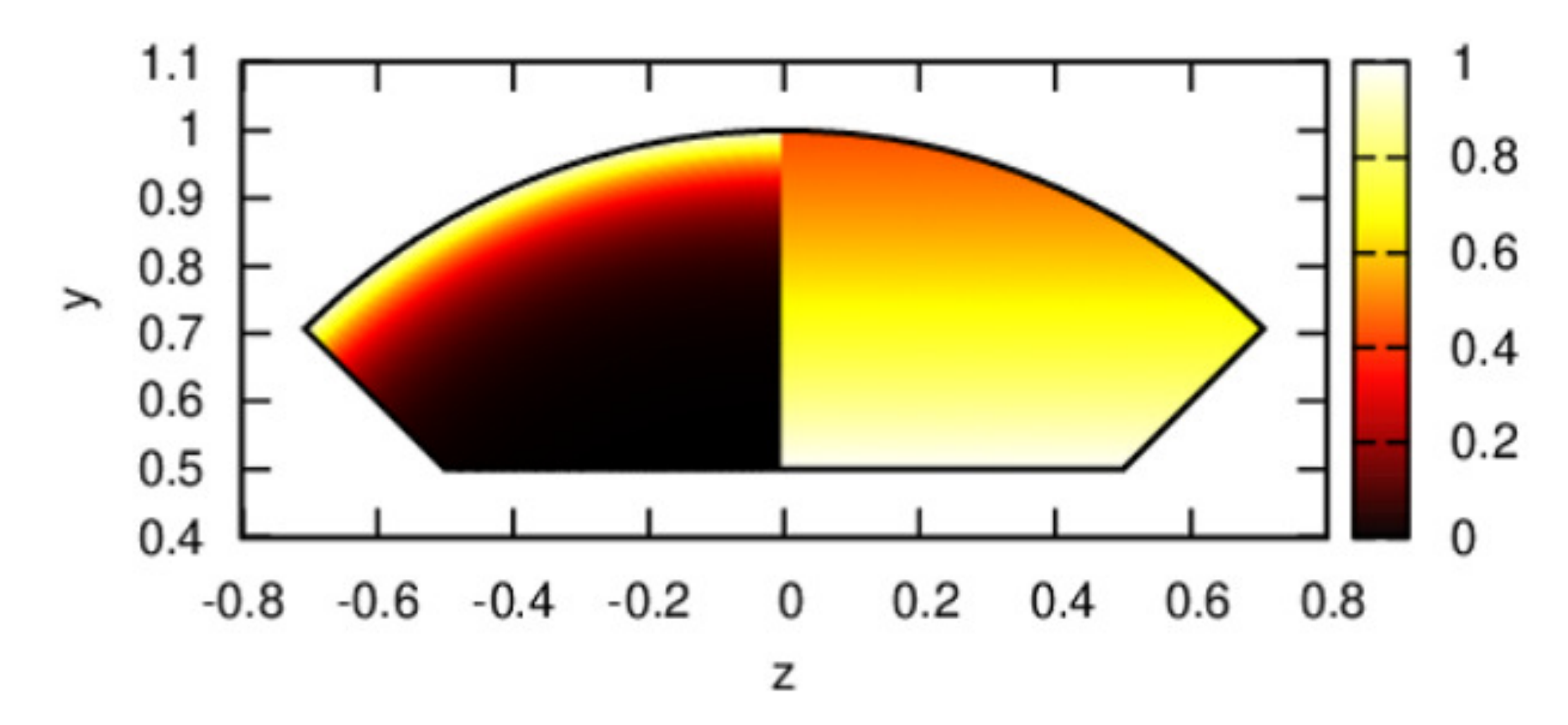}
 \caption{The model base flow introduced in section 2.4. $\alpha=15$, $g=1$.  The flow profile is symmetric with respect to $z=0$ axis. In the left half of the domain the normalised base shear $\overline{u}/u_0$ is plotted, while the right half is the base density $\overline{\rho}$. The applied magnetic field is uniform, and the fluid is isothermal. 
}
\end{figure}
In sections 3 and 4 we will derive three eigenvalue bounds. 
Since it is desirable to consider a specific flow to study the nature of the bounds, here we introduce a model flow.
\textcolor{black}{
The model is designed to be as simple as possible in its representation.
The only conditions we applied in the design are that the fluid flows through a duct of non-trivial cross section and that the stratification and shear directions should not be aligned. The simplicity of the model is advantageous for the purpose of reproducing the results of this paper.
Recall that our goal is not to solve any specific application problem, but to derive generic properties that hold for a wide range of flows.
}

Consider a flow through a duct of a cross-section $\Omega$, which is bounded by $z=1/2, r=1, \theta=\pi/4$ and $\theta=3\pi/4$; see figure 1.
Here the cylindrical coordinates $(r,\theta)$ in the $y,z$ plane are defined by $z=r\cos \theta, y=r\sin \theta$.
We assume that the base velocity profile develops a boundary layer near $r=1$ so that
\begin{eqnarray}
\overline{u}=u_0\frac{e^{\alpha (r-1)}-e^{-\alpha/2}}{1-e^{-\alpha/2}}.
\end{eqnarray}
The parameter $\alpha$ controls the strength of the shear in the boundary layer, and $u_0$ specifies the velocity range (note that $\overline{u}\in [0,u_0]$). We use $u_0=0.5$ and $\alpha=15$ throughout the paper.

The fluid is assumed to be isothermal, and subjected to a downward uniform gravity field $G=-gz$ and a uniform streamwise magnetic field. Here $g$ is a positive constant representing the strength of the gravity.
Without loss of generality we can normalise the flow so that the sound wave speed becomes unity. 
The density profile $\overline{\rho}=e^{\gamma g (\frac{1}{2}-z)}$
can be found by solving the magneto-static condition with $\overline{\rho}|_{z=\frac{1}{2}}=1$.

The square of the Alfv\'en wave speed is inversely proportional to the density as
\begin{eqnarray}
a^2=\frac{2}{\gamma \beta_0}e^{\gamma g (z-\frac{1}{2})}.\label{modelAsq}
\end{eqnarray}
Here $\beta_0$ is the plasma beta (the ratio of the kinematic pressure to the magnetic pressure) at $z=\frac{1}{2}$. 
The cusp wave speed can be computed by (\ref{defcusp}) using (\ref{modelAsq}) and $s^2=1$. Due to the normalisation we applied, when $u_0$ is less than unity the base velocity field becomes subsonic everywhere in $\Omega$. 

The buoyancy frequency can be found from (\ref{N12}):
\begin{eqnarray}
N_1^2=0,\qquad N_2^2=g^2(\gamma-1),\qquad N_{12}^2=0.
\end{eqnarray}
Hereafter when $N_1^2+N_2^2>0$ everywhere in $\Omega$, we say that the base flow is convectively stably stratified.
For the standard choice $\gamma=5/3$ we adopt, clearly the model flow satisfies this condition.
The parameters in the model flow are $g$ and $\beta_0$. The parameter $g$ can be used to control the strength of the stratification, while the inverse of $\beta_0$ represents the magnitude of the square of the applied magnetic field (thus small $\beta_0$ implies that the applied magnetic field is strong, and vice versa).

\textcolor{black}{
In this paper, numerical stability analysis of the model flow is not dealt with because the numerical study of non-planar flows is far beyond the scope of this study. 
Here, we shall briefly explain why a such numerical computation is so challenging. 
First of all, it is well-known that the eigenvalue computation is quite unstable for inviscid problems.
The convergence of the iterative methods is not very efficient because of a bad condition number, while
in the direct methods an accurate computation is difficult due to the singularities that appear in the neutral solutions. 
Even for unstable eigenmodes, the numerical growth rates are typically not so large, and very sharp structures occur in the corresponding eigenmodes. This is particularly problematic in the non-planar problems because the very high resolution required makes the matrix size huge. 
For the planar cases, the shooting method is a prescription for this problem but of course it is not applicable for the non-planar problems.
}

\textcolor{black}{
To see how the bounds and eigenvalues compare, it is sufficient to use planar flows, as we will see in section 5. Even if we could perform non-planar calculations, we would not get much useful information, since our interest is not in studying a particular flow.
}
 


\section{Semicircle type theorems}
If the flow is assumed to be independent of $y$ and the gravitational field is uniform, our flow configuration reduces to that considered in Adam (1978b), where the equation for $\widetilde{q}$ is used to derive a semicircle theorem for convectively stable or neutrally stable flows. 
Unfortunately, it turns out that the same strategy cannot work for the generalised flow configuration considered in this paper. 
The issue here can nevertheless be resolved by simply going back to the earlier work by Eckart (1963), who used the displacement form of the equations. 
We shall see that there is a close relationship between the well-known energy principle theory for the displacement and the semicircle theorem, which even allows for much refinement of the eigenvalue bound.


\subsection{Usual semicircle theorem}
Let us begin our analysis by briefly reviewing the energy principle theory that forms the basis of the derivation of the semicircle theorem. 
The energy balance in terms of the Lagrange description can be analysed by first taking the inner product of (\ref{xiforce}) and the complex conjugate of the displacement vector, and then integrating it by parts over the domain $\Omega$.
\begin{eqnarray}
c^2\langle Q \rangle -2c\langle \overline{u}Q \rangle -\delta W=0.\label{quadc}
\end{eqnarray}
Here
\begin{eqnarray}
Q\equiv \overline{\rho}\, k^2(|\xi|^2+|\eta|^2+|\zeta|^2) 
\end{eqnarray}
is positive definite, and the angle brackets represent the integration over the domain. 
The first, second, and third terms on the left side of (\ref{quadc}) are referred to as the kinematic energy, gyroscopic term, and potential energy, respectively.
The last term $\delta W$ is of course from the force operator $\mathbf{F}$.
Bernstein et al. (1958) showed for the static case ($\overline{u}$=0) that the base state is stable if and only if $\delta W$ is positive definite, while in the presence of the shear ($\overline{u}\neq$0) the positiveness only guarantees the sufficient condition of the stability (Frieman \& Rotenberg 1960). Those properties immediately follow by solving the quadratic equation (\ref{quadc}) for $c$.

The terms in (\ref{quadc}) can be rearranged to obtain the following form that is more suitable to obtain the semicircle theorem:
\begin{eqnarray}
\langle U^2Q \rangle=\langle \mathcal{L} \rangle.\label{energy}
\end{eqnarray}
The integral appeared on the right side is $\langle  \mathcal{L}\rangle=\delta W+\langle\overline{u}^2Q \rangle$ and the explicit form of $\mathcal{L}$ can be found by (\ref{1rho}), (\ref{2q}), and (\ref{momxi}) as
\begin{eqnarray}
\mathcal{L}&=&
\overline{\rho}s^2|ik \xi+\eta_y+\zeta_z|^2
+\overline{\rho}a^2\{k^2   (|\eta|^2+ |\zeta|^2)+|\eta_y+\zeta_z|^2\}\nonumber \\
&&~~~+\overline{\rho}\{(G_y\eta+G_z\zeta)^*(ik \xi+\eta_y+\zeta_z)+(G_y\eta+G_z\zeta)(ik \xi+\eta_y+\zeta_z)^*\}\nonumber \\
&&~~~+(G_y\eta+G_z\zeta)^*(\overline{\rho}_y\eta+\overline{\rho}_z\zeta) \nonumber \\
&=&
\overline{\rho}s^2|ik \xi+\eta_y+\zeta_z+s^{-2}(G_y\eta+G_z\zeta)|^2
+\overline{\rho}a^2|\eta_y+\zeta_z|^2\nonumber \\
&&~~~+\overline{\rho}\{
(N_1^2+k^2a^2)|\eta|^2
+(N_2^2+k^2a^2)|\zeta|^2
+N_{12}^2(\zeta^*\eta+\eta^*\zeta)
\}.
\label{LLL}
\end{eqnarray}
Note that $Q$ and $\mathcal{L}$ are real-valued functions. 
Thus the real and imaginary parts of equation (\ref{energy}) becomes
\begin{eqnarray}
\langle \{(\overline{u}-c_r)^2-c_i^2\} Q \rangle=\langle \mathcal{L} \rangle,\label{Qreal}\\
-2c_i\langle (\overline{u}-c_r) Q \rangle=0,\label{Qimag}
\end{eqnarray}
respectively.

First we show that the semicircle theorem holds for unstable modes ($c_i\neq 0$) as long as $\langle \mathcal{L}\rangle \geq 0$ is satisfied, and then check when the latter condition is satisfied.
As usual for Howard's type theory, we combine (\ref{Qreal}) with the obvious inequality
\begin{eqnarray}
\left \langle  \left (\overline{u}-\min_{\Omega}\overline{u}\right )\left (\overline{u}-\max_{\Omega}\overline{u}\right )Q \right\rangle
=
\langle 
\{\overline{u}^2-2\overline{u}_+c_r+\overline{u}_+^2-\overline{u}_-^2\}Q
\rangle
 \leq 0,\label{ineq2}
\end{eqnarray} 
where
\begin{eqnarray}
\overline{u}_\pm \equiv \frac{1}{2} \left (\max_{\Omega}\overline{u}\pm \min_{\Omega}\overline{u}\right ).
\end{eqnarray}
Noting that (\ref{Qimag}) now implies $\langle \overline{u}Q \rangle=c_r \langle Q \rangle$, we have
\begin{eqnarray}
\langle \{(c_r-\overline{u}_+)^2+c_i^2\}Q \rangle \leq  \langle \overline{u}_-^2Q\rangle,
\end{eqnarray}
which is nothing but the semicircle theorem $(c_r-\overline{u}_+)^2+c_i^2 \leq \overline{u}_-^2$ ensuring that in the complex plane the unstable eigenvalue $c$ lies inside, or on, the semicircle in the upper half-plane, whose centre and radius are $\overline{u}_+$ and $\overline{u}_-$, respectively.


There is a simple  condition to guarantee $\langle \mathcal{L}\rangle \geq 0$.
The terms in the curly bracket in last line of (\ref{LLL}) can be written in a quadratic form 
\begin{eqnarray}
[\eta,\zeta]
\left [
\begin{array}{cc}
N_1^2+k^2a^2 & N_{12}^2\\
N_{12}^2 & N_2^2+k^2a^2
\end{array}
\right ]
\left [
\begin{array}{c}
\eta\\
 \zeta
\end{array}
\right ].
 \label{NNNN}
\end{eqnarray}
As well-known, the definiteness of the Hermitian quadratic form can be found by the eigenvalues of the matrix; hence $\mathcal{L} \geq 0 $ must hold if the eigenvalues of the above real symmetric matrix are non-negative for all points $y,z$.
The eigenvalues are found as $k^2a^2$ and \textcolor{black}{$k^2a^2+N_1^2+N_2^2$}, noting the identity $N_{12}^4=N_1^2N_2^2$ (see (\ref{magstat}) and (\ref{N12})).
Therefore, the only caveat in using the semicircle theorem is that $k^2a^2+(N_1^2+N_2^2)$ is positive everywhere.
In particular, when the flow is not convectively unstably stratified (i.e. if $N_1^2+N_2^2\geq 0$ is satisfied at all points in $\Omega$), the semicircle theorem holds for any wavenumber $k$.


The result here extends the semicircle theorem for quite general flow configurations including those studied in Eckart (1963) and Adam (1978b). 
This new result is already remarkable, but note that the derivation above might be not so surprising because for ideal fluids the force operator $\mathbf{F}$ in (\ref{xiforce}) typically possesses the self-adjoint property, which is all we need to show that
$\mathcal{L}$ in (\ref{energy}) is purely real. 
The more substantive findings of this paper will be shown in the next section, where we shall see that the energy principle equation allows us to improve the semicircle bound.

\subsection{Inner envelope theorem}
The main tool to be used to improve the eigenvalue bound is the simple identity
\begin{eqnarray}
\langle\{\overline{u}^2-2\overline{u}c_r+c_r^2-c_i^2\}Q \rangle
=
\langle\{
(\overline{u}-r_c)^2
+2\overline{u}(r_c-c_r)
+c_r^2-r_c^2
-c_i^2\}Q \rangle \nonumber \\
=
\langle\{
(\overline{u}-r_c)^2
-(c_r-r_c)^2
-c_i^2\}Q \rangle \label{envmain}
\end{eqnarray}
that holds for an arbitrary real number $r_c$.
Here (\ref{Qimag}) is used in the second equality.
Equation (\ref{Qreal}) can thus be transformed into
\begin{eqnarray}
\langle \{(c_r-r_c)^2+c_i^2\} Q \rangle=\langle (\overline{u}-r_c)^2 Q \rangle-\langle \mathcal{L} \rangle.
\label{envsemi}
\end{eqnarray}
If we can find a positive number $R(r_c)$ such that 
\begin{eqnarray}
\langle (\overline{u}-r_c)^2 Q \rangle-\langle \mathcal{L} \rangle\leq R^2\langle Q\rangle,\label{envsemi2}
\end{eqnarray}
we can establish a semicircle theorem with the centre $r_c$ and the radius $R(r_c)$. 
The theory works for any real number $r_c$, and hence the net eigenvalue bound in the complex plane can be obtained by drawing the semicircles changing $r_c$.

Here we shall make a few remarks on the relationship between the new eigenvalue bound and the usual semicircle bound.
First, if $\langle \mathcal{L}\rangle= 0$, namely if the flow is unstratified, incompressible and non-magnetised, the best bound found by (\ref{envsemi}) coincides with the usual semicircle. This can be easily checked by noticing $\min_{\Omega}(\overline{u}-r_c)=\overline{u}_+-\overline{u}_--r_c$ and $\max_{\Omega}(\overline{u}-r_c)=\overline{u}_++\overline{u}_--r_c$. Then the inequality
$\langle (\overline{u}-r_c)^2 Q \rangle\leq (|\overline{u}_+-r_c|+|\overline{u}_-|)^2\langle Q \rangle$ implies
  that the semicircles for any $r_c$ contain the usual semicircle.
Second, if $\langle \mathcal{L}\rangle> 0$ the usual semicircle can be found by the choice $r_c=\overline{u}_+$ that enable us to use $R=\overline{u}_-$ from the trivial inequality $\langle (\overline{u}-\overline{u}_+)^2 Q \rangle-\langle \mathcal{L} \rangle < \langle (\overline{u}-\overline{u}_+)^2 Q \rangle \leq \overline{u}_-^2 \langle Q\rangle$.
One may notice that this choice of $R$ might be not optimal because the property of $\langle \mathcal{L}\rangle$ is not used except for its positiveness. 
In fact, the reason why the envelope semicircle theorem below gives a better bound is that it uses that property more effectively.



Of course, in order to draw a bound in the complex plane, we must somehow find $R(r_c)$ that satisfies (\ref{envsemi2}).
The best possible eigenvalue bound could be found by seeking the optimum value of $R^2$ for all possible displacement functions:
\begin{eqnarray}
R^2\equiv \max_{\xi,\eta,\zeta} \frac{\langle (\overline{u}-r_c)^2Q-\mathcal{L}\rangle}{\langle Q \rangle},\label{maxQL}
\end{eqnarray}
where the value of $R^2$ may be able to find by treating the Euler-Lagrange equations as an eigenvalue problem for eigenvalue $R^2$ (see Appendix B).
However, this method is not practically very useful, because the computational effort to solve this eigenvalue problem is comparable to that for the stability problem. 
Besides we may need to check if the stationary point found by the Euler-Lagrange equations is really a maximum, considering the second variational problem.

There is an analytical way to estimate $R$ that satisfies (\ref{envsemi2}).
Since ease of computation is an advantage in applying the bounds to practical problems, in this paper we focus on this method.
We first note that introducing an arbitrary positive function $\sigma (y,z)$, the integrand $\mathcal{L}$ seen in (\ref{LLL}) can be rewritten in the form
\begin{eqnarray}
\mathcal{L}&=&\overline{\rho}\left \{\sigma \left |\eta_y+\zeta_z+\frac{ik s^2\xi+G_y\eta+G_z\zeta}{\sigma}\right |^2+(s^2+a^2-\sigma)|\eta_y+\zeta_z|^2\right \}\nonumber \\
&&+ \overline{\rho}\boldsymbol{\xi}^{\dagger}(k^2\mathbb{L}_2+k \mathbb{L}_1+\mathbb{L}_0)\boldsymbol{\xi},\label{newL}
\end{eqnarray}
where $\boldsymbol{\xi}$ is the transpose of $[i\xi,\eta,\zeta]$, and
\begin{eqnarray*}
\mathbb{L}_2=
\left [
\begin{array}{ccc}
s^2(1-\frac{s^2}{\sigma}) & 0 & 0 \\
0& a^2 & 0 \\
0& 0 & a^2
\end{array}
\right ],
~~
\mathbb{L}_1=(1-\frac{s^2}{\sigma})
\left [
\begin{array}{ccc}
0 & G_y & G_z \\
G_y & 0 & 0 \\
G_z & 0 & 0
\end{array}
\right ],~~\\
\mathbb{L}_0=
\left [
\begin{array}{ccc}
0 & 0 &0 \\
0& \frac{G_y\overline{\rho}_y}{\overline{\rho}}-\frac{G_y^2}{\sigma}&  \frac{G_z\overline{\rho}_y}{\overline{\rho}}-\frac{G_yG_z}{\sigma} \\
0& \frac{G_z\overline{\rho}_y}{\overline{\rho}}-\frac{G_yG_z}{\sigma} & \frac{G_z\overline{\rho}_z}{\overline{\rho}}-\frac{G_z^2}{\sigma}
\end{array}
\right ],\hspace{20mm}
\end{eqnarray*}
are real symmetric matrices. 
The eigenvalues of $\mathbb{L}_1$ and $\mathbb{L}_0$ can be worked out analytically as
\textcolor{black}{
\begin{subequations}\label{eigen12}
\begin{eqnarray}
\left \{0,(1-\frac{s^2}{\sigma})\sqrt{G_y^2+G_z^2}\}, -(1-\frac{s^2}{\sigma})\sqrt{G_y^2+G_z^2} \right \},\\
\left \{0,0,\frac{G_y\overline{\rho}_y+G_z\overline{\rho}_z}{\overline{\rho}}-\frac{G_y^2+G_z^2}{\sigma}\right \},
\end{eqnarray}
\end{subequations}
}
respectively.
The new form of $\mathcal{L}$ suggests that as long as $0< \sigma \leq s^2+a^2$ holds everywhere, the terms in the curly bracket in (\ref{newL}) are positive definite, and thus we can drop them in the estimation of the radius $R$.


Now, suppose at each $y,z$ the largest eigenvalues of the matrices $(\overline{u}-r_c)^2\mathbb{I}-\mathbb{L}_2$, $-\mathbb{L}_1$, $-\mathbb{L}_0$ are calculated ($\mathbb{I}$ is a 3 by 3 identity matrix); we denote them as $\lambda_2(y,z)$, $\lambda_1(y,z)$, $\lambda_0(y,z)$, respectively. Then if $0<\sigma \leq s^2+a^2$ everywhere, from (\ref{envsemi}) and (\ref{newL}) we can deduce
\begin{eqnarray}
\langle \{(c_r-r_c)^2+c_i^2\}Q \rangle \leq k^2 \langle \overline{\rho}\boldsymbol{\xi}^{\dagger}\{(\overline{u}-r_c)^2\mathbb{I}-\mathbb{L}_2-\lambda_2\mathbb{I}\}\boldsymbol{\xi} \rangle \nonumber \\
+k \langle \overline{\rho}\boldsymbol{\xi}^{\dagger}\{-\mathbb{L}_1-\lambda_1\mathbb{I}\}\boldsymbol{\xi} \rangle
+\langle \overline{\rho}\boldsymbol{\xi}^{\dagger}\{-\mathbb{L}_0-\lambda_0\mathbb{I}\}\boldsymbol{\xi} \rangle \nonumber \\
+\langle (\lambda_2+k^{-1}\lambda_1+k^{-2}\lambda_0) Q\rangle \nonumber\\
\leq \langle (\lambda_2+k^{-1}\lambda_1+k^{-2}\lambda_0) Q\rangle \leq R^2\langle Q \rangle,
\end{eqnarray}
where we have defined the radius $R$ so that
\begin{eqnarray}
R^2\equiv \max_{\Omega}(\lambda_2+k^{-1}\lambda_1+k^{-2}\lambda_0),
\end{eqnarray}
where $R^2$ is of course dependent on our choice of the function $\sigma$. Optimising this function so that we have the tightest possible bound (see Appendix C), we have the following theorem.

\textcolor{black}{
\begin{thm}
The unstable complex phase speed $c=c_r+ic_i$ of (\ref{full_linear}) is bounded so that $(c_r-r_c)^2+c_i^2\leq \{R(r_c)\}^2$ for any real number $r_c$. Here the radius $R$ is determined by $R^2=\max_{\Omega}\lambda(y,z)$, where
\begin{eqnarray*}
\lambda(y,z)=
\left \{
\begin{array}{c}
(\overline{u}-r_c)^2-c_T^2\max(1-\frac{k_h}{k},0)
~~~~~~~~~~~~~~~~~~\text{if}~~~N_1^2+N_2^2\geq 0,\\
(\overline{u}-r_c)^2-c_T^2(1-\frac{k_h}{k})+\max(\frac{N_{1a}^2+N_{2a}^2}{k_h^2},0)\max(1-\frac{k_h}{k},0)~\\
~~~~~~~~~~~~~~~~~~~~~~~~~~~~~~~~~~~-\min(\frac{N_{1a}^2+N_{2a}^2}{k^2},0)~~~~\text{otherwise}.
\end{array}
\right .
\end{eqnarray*}
\end{thm}
}

Here we have defined the magnetic buoyancy frequencies
\begin{eqnarray}
N_{1a}^2\equiv \frac{G_y\overline{\rho}_y}{\overline{\rho}}-\frac{G_y^2}{s^2+a^2},\qquad
N_{2a}^2\equiv \frac{G_z\overline{\rho}_z}{\overline{\rho}}-\frac{G_z^2}{s^2+a^2},
\end{eqnarray}
the local cusp (tube) wave speed
\begin{eqnarray}
c_T\equiv \sqrt{\frac{s^2a^2}{s^2+a^2}},\label{defcusp}
\end{eqnarray}
and the wavenumber $k_h\equiv h^{-1}$ associated with the density scale height
\begin{eqnarray}
h\equiv \frac{s^2}{\sqrt{G_y^2+G_z^2}}.
\end{eqnarray}

A notable property of this new bound is that it works without any caveat; recall that the usual semicircle theorem works only when $\langle \mathcal{L}\rangle\geq 0$.
Therefore, from the theorem, an eigenvalue bound can be deduced even when the flow is strongly unstably stratified.
While when the flow is not convectively unstably stratified $(N_1^2+N_2^2\geq 0)$ everywhere in $\Omega$, the bound should become tighter than the usual semicircle bound because of the term $c_T^2\max(1-\frac{k_h}{k},0)\geq 0$ in the function $\lambda(y,z)$ (recall that if this term is absent, we have the usual semicircle bound). 
Due to the stabilisation, the radius $R(r_c)$ can become zero at some $r_c$, in which case we can establish the stability of the flow. This is another feature that has not been seen in the usual semicircle theorem.

\begin{figure}
\hspace{-35mm}(a)\\
\centering
(b)\includegraphics[scale=1.]{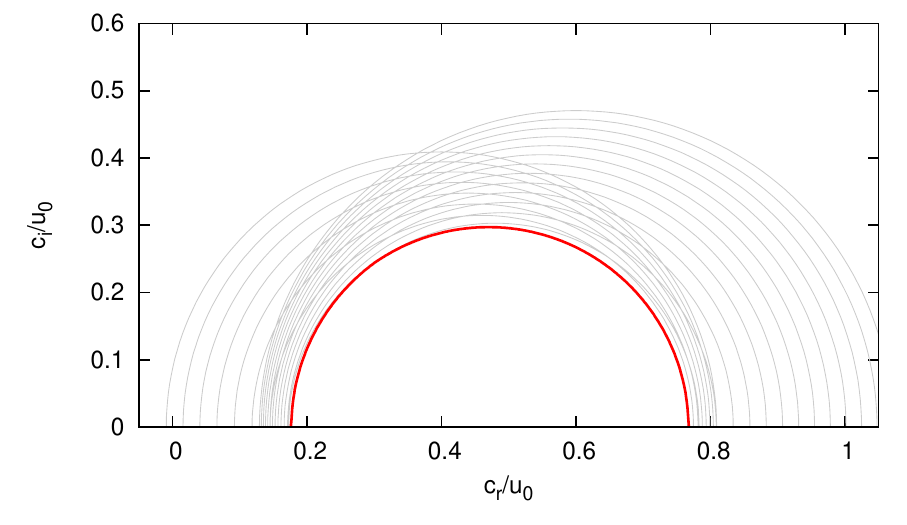}\\ 
(c)\includegraphics[scale=1.]{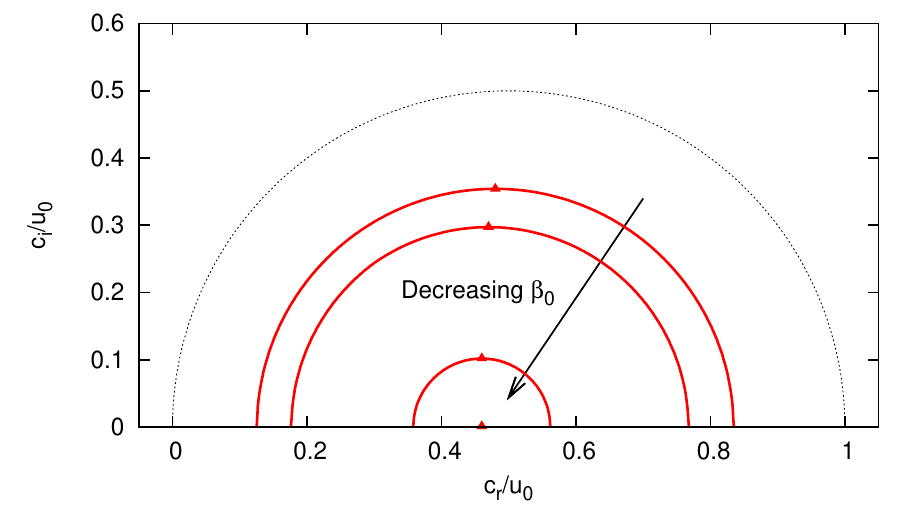} \\
\includegraphics[scale=1.]{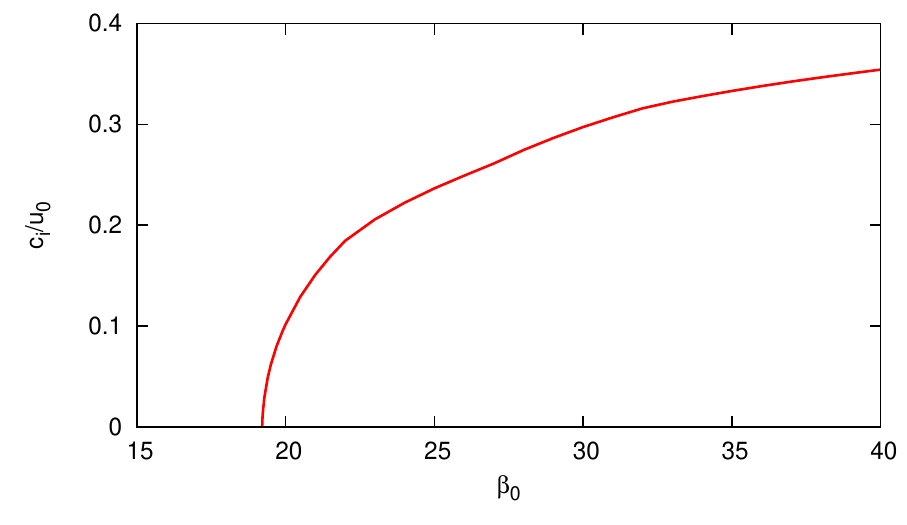}
\caption{
The eigenvalue bounds by theorem 1 for the model flow defined in section 2.4. $(k,u_0,g)=(10, 0.5, 1)$.
(a): The thin grey curves are the semicircles for various choices of $r_c$ for $\beta_0=30$, and the red thick curve is their inner envelope (i.e. the net eigenvalue bound).
(b): The comparison of the inner envelope bounds for $\beta_0=19.2,20,30,40$. The triangles indicate the maximum $c_i$ attained in each bound. 
The thin dotted circle is the usual semicircle bound.
(c): The variation of the the maximum $c_i$ predicted by the inner envelope bound. 
%
%
}
\label{fig:longwave}
\end{figure}

Here we use the model flow defined in section 2.4 to see how such a bound/stability condition can be deduced from the above theorem.
Figure 2a demonstrates the derivation of the net eigenvalue bound for $\beta_0=30$. 
The thin grey circles are the semicircles with various choices of $r_c$.
We can easily compute the radius $R(r_c)$ of them by maximising the quantity $\lambda$ over $\Omega$
Then, since the eigenvalues must lie within the semicircle for any $r_c$, the net bound can be found by taking the \textit{inner envelope} of the semicircles, as shown by the thick red curve in the figure. 
Recall that the model flow is convectively stably stratified. Thus, as remarked above, the inner envelope bound should become tighter than the semicircle theorem. This can be confirmed in figure 2b, 
where the inner envelope bounds for $\beta_0=19.2, 20,30,40$ are shown by the red curves.
The red triangles on the bounds show the maximum $c_i$ attained by the corresponding bound. The variation of that maximum with respect to $\beta_0$ is shown in figure 2c.
As the $\beta_0$ decreases, the bound becomes tighter, which indicates the stabilising effect of the applied magnetic field. 
We can show why this happens using the explicit form of the bound.
As the applied magnetic field becomes stronger, the cusp wave speed $c_T$ increases 
, and hence the aforementioned term $c_T^2\max(1-\frac{k_h}{k},0)\geq 0$ representing the stabilisation effect becomes more dominant.
%
It is clear from figure 2c that when $\beta_0$ is smaller than $19.2$ the inner envelope bound predicts the stability of the flow.


A similar stabilisation effect by the applied magnetic field was reported in some previous unstratified incompressible planar flow studies, and they are in fact included in our result.
The incompressible limit $s\rightarrow \infty$ implies $k_h\rightarrow 0$ and $c_T\rightarrow a$.
Therefore, we simply have $R^2=\max_{\Omega}\{(\overline{u}-r_c)^2-a^2\}$, where the term proportional to the square of the Alfv\'en wave speed essentially describes the stabilisation effect for example found by Howard \& Gupta (1962). 
More specifically, we can show that the incompressible, unstratified, and planar version of our result contains the two bounds found by Hughes \& Tobias (2001). If we set $r_c=0$, our bound becomes $c_r^2+c_i^2\leq \max_{\Omega} (\overline{u}^2-a^2)$, which is identical to (3.10) in Hughes \& Tobias (2001). 
While if we set $r_c=\overline{u}_+$ the bound becomes $(c_r-\overline{u}_+)^2+c_i^2\leq \max_{\Omega} ((\overline{u}-\overline{u}_+)^2-a^2)$, from which we can deduce $(c_r-\overline{u}_+)^2+c_i^2\leq \overline{u}_-^2-\min_{\Omega} a^2$, namely (3.14) of Hughes \& Tobias (2001).
Gupta (1992) treated the incompressible version of our non-planar system but as remarked by Cally (2000) there are some errors. 
The corrected bound is essentially the second bound in Hughes \& Tobias (2001).

%

%
%

A crucial difference between the above incompressible results and our compressible stratified result is that whether the stabilisation occurs or not depends on the wavenumber due to the term proportional to $k_h/k$ in $\lambda(y,z)$. 
%
For the convectively stable flows, 
in the limit of $k_h/k\rightarrow 0$ the bound becomes tightest and Cally's stability condition derived for unstratified flows can be recovered; \textit{the flow is stable if there exists a Galilean frame in which the flow nowhere exceeds the cusp speed}. 
(This can be easily checked by noting our expression of the radius becomes $R^2=\max_{\Omega}\{(\overline{u}-r_c)^2-c_T^2\}$ in the limit; Cally (2000) considered planar flows but from the result here clearly his stability condition can be carried over to non-planar flows.)
On the other hand, the stabilisation does not occur at all if $k$ is smaller than $k_h$ -- namely \textit{when the wavelength is larger than the density scale height, the bound becomes the usual semicircle. }
As an example, consider the model flow used in figure 2 again. The wavenumber associated with the density scale height is $k_h=g=1$. The reason why we had effective magnetic stabilisation in figure 2 is that the wavenumber $k=10$ we used is much larger than $k_h$. 
For $k\leq 1$, the ratio $k_h/k$ is greater than unity and thus the net bound is merely the usual semicircle whatever the value of $\beta_0$ is. 
Theorem 1 implies that somewhat counter-intuitively, the presence of the stable stratification may hinder the stabilisation effect by the applied magnetic field. 


The behaviour of the bound for the unstably convectively stratified flows is more complicated, but at least it is easy to check that when there is no velocity shear the bound is consistent with the sufficient condition of stability by Newcomb (1961) (note that Newcomb (1961) also deduced the necessary condition of stability). 
To check this, we may set $(\overline{u}-r_c)^2=0$ in our bound. The sign of $\lambda(y,z)$ can be easily found in Table 1.
If $0\leq N_1^2+N_2^2$, clearly, $\lambda(y,z)\leq 0$ and thus $R^2\leq 0$ for all $k$.
This sufficient condition of stability is precisely what was shown by Newcomb (1961), though it is now extended for the general conservative force field.
In section 5, we will examine the bound for a specific convectively unstable shear flow.




Finally. we briefly comment on the stability condition derived in
Adam (1978a), who also used the energy principle theory assuming a planar base field depending only on $z$. 
His method is essentially based on the minimisation problem of $\delta W=\langle \mathcal{L}-(\overline{u}-r_c)^2Q\rangle$. If it is positive for all possible displacements, as remarked at the beginning of this section, the flow must be stable (see Frieman \& Rotenberg (1960)).
The displacement vector that gives the stationary value of $\delta W$ could be found by the Euler-Lagrange equations.  Adam (1978a) used those equations to eliminate the two displacement components from the stationary value of $\delta W$ and make it a function of only one remaining displacement component.
However, there is a flaw in this key step towards his stability condition.
His Euler-Lagrange equations almost always do not have solution other than the trivial one $(\xi=\eta=\zeta=0)$. 
Thus the stationary value of $\delta W$ is merely zero (that makes sense because $\delta W$ can be written in a quadratic form), and thus the sign of it does not supply a meaningful stability condition. 
Moreover, the second variational problem considered in that paper treats variations of only two displacement components, and the consideration of it does not guarantee that the stationary point is minimum for all possible displacements.

Here one might notice that Adam's method has some similarity to our first method above. 
The optimisation problem (\ref{maxQL}) resembles what was considered in Adam (1978a), but the crucial difference is that in our method $\delta W$ is normalised by $\langle Q \rangle$, which is essentially the norm of the displacement.
This normalisation makes the stationary value to be non-trivial and produces a Lagrange multiplier $R^2$ in our Euler-Lagrange equations.
In other words, the error in Adam (1978a) essentially comes from the fact that he assumed the existence of a non-trivial Euler-Lagrange solution associated with the eigenvalue $R^2=0$, which is almost always impossible.


The derivation of the stability conditions by Hughes \& Tobias (2004) is based on the theory by Adam (1978a) and hence not correct as well.
Therefore, there is no simple stability condition available for the flows influenced by shear, stratification, compressibility, and magnetic field, except for that deduced in theorem 1.



\section{Generalisation of the Miles-Howard theory, the semi-ellipse type theorem, and Cally's eigenvalue bound}


Our primal aim of this section is to generalise the Miles-Howard theory.
There are three effects to be included; the compressible, magnetohydrodynamic, and non-planar effects. 
The conclusion to be deduced in section 4.1 is that unfortunately two of those effects prevent the derivation of a stability condition.
However, the equations obtained in the course of the analysis are nevertheless useful, because they help us to derive two other types of eigenvalue bounds.

The first bound is somewhat similar to the semi-ellipse theorem by Kockar \& Jain (1979) derived for planar flows, although it is important to remark here that their derivation is not applicable when the base flow depends on two variables. 
The alternative idea of using the Lagrangian displacement is motivated by Fung (1986), but as remarked in section 1 his analysis was not complete.
The second bound is similar to that derived in Cally (1983). Again, we note that the method used in that paper cannot be used for non-planar flows, because it is based on the theory by Warren (1970) for certain second-order ordinary differential equations. Thus our analyses in sections 4.2 and 4.3 are not merely a minor extension of the previous works but involve new methodologies.




\subsection{The Miles-Howard theory for the generalised flow}

Following Miles (1961) and Howard (1961), we rescale the perturbation quantities by $U^{1/2}$.
The governing equations for the rescaled displacement $\varphi=U^{1/2}\xi$, $\phi=U^{1/2}\eta$, $\psi=U^{1/2}\zeta$ can be found from (\ref{1rho}), (\ref{2q}) and (\ref{momxi}) as
\begin{subequations}
\begin{eqnarray}
U^{1/2}\widetilde{q}=-\overline{\rho}a^2\widehat{\kappa}-(s^2\overline{\rho}(ik\varphi+\widehat{\kappa})+\overline{\rho}\widehat{\mathcal{G}}),\label{qqeq}\\
\overline{\rho}k^2 U^2 \varphi
=-ik (s^2\overline{\rho}(ik\varphi+\widehat{\kappa})+\overline{\rho}\widehat{\mathcal{G}})\label{vareq},\\
k^2\overline{\rho}U^{-1}(U^2-a^2) \phi=U^{-1/2}\widetilde{q}_y
+U^{-1}G_y(\overline{\rho}_y\phi+ \overline{\rho}_z\psi+\overline{\rho}(ik\varphi+\widehat{\kappa})),\label{phieq}\\
k^2\overline{\rho}U^{-1}(U^2-a^2) \psi=U^{-1/2}\widetilde{q}_z
+U^{-1}G_z( \overline{\rho}_y\phi+\overline{\rho}_z\psi+\overline{\rho}(ik\varphi+\widehat{\kappa})).\label{psieq}
\end{eqnarray}
\end{subequations}
For the sake of simplicity here we have denoted as
\begin{subequations}\label{defkappa}
\begin{eqnarray}
\widehat{\mathcal{G}}=G_y\phi+G_z\psi,\\
\widehat{\kappa}=U^{1/2}(\eta_y+\zeta_z)=
\widehat{\kappa}_1-U^{-1}\widehat{\kappa}_2,\\
\widehat{\kappa}_1=\phi_y+\psi_z,\qquad \widehat{\kappa}_2=\frac{1}{2}(U_y\phi+U_z\psi).
\end{eqnarray}
\end{subequations}

Next we multiply $\phi^*$ and $\psi^*$ to (\ref{phieq}) and (\ref{psieq}) respectively, and then add them together. Integrating the resultant equation by parts over the domain, and taking the imaginary part, we can find that the unstable mode must satisfy (see Appendix D).
\begin{eqnarray}
0=\left \langle Q_1+\frac{\widehat{f}_b}{|U|^2}-\frac{\overline{\rho}|\widehat{\kappa}_2|^2}{|U|^2}\left (\frac{s^2}{|U|^2+s^2}+\frac{a^2}{|U|^2}\right )
\right \rangle.\label{compQ1}
\end{eqnarray}
Here
\begin{eqnarray}
Q_1&\equiv&\overline{\rho}\left (1+\frac{a^2}{|U|^2} \right )k^2(|\phi|^2+|\psi|^2)\nonumber \\
&&+\overline{\rho}s^2\frac{|U|^2+s^2}{|U^2-s^2|^2}\left |\widehat{\kappa}_1-\frac{2U_r}{|U|^2+s^2}\widehat{\kappa}_2+\frac{\widehat{\mathcal{G}}}{s^2}\right |^2
+\overline{\rho}\frac{a^2}{|U|^2}\left |\widehat{\kappa}-\frac{2U_r}{|U|^2}\widehat{\kappa}_2\right |^2\label{defQ1}
\end{eqnarray}
is a real positive function ($U_r=\overline{u}-c_r$ is the real part of $U$) and 
\begin{eqnarray}
\widehat{f}_b\equiv \frac{G_y}{\overline{\rho}_y}|\overline{\rho}_y\phi+\overline{\rho}_z\psi|^2-\frac{\overline{\rho}}{s^2}|\widehat{\mathcal{G}}|^2\label{deffb}
\end{eqnarray}
denotes the buoyancy related terms.
\textcolor{black}{The equation (\ref{compQ1}) plays the central role in the Miles-Howard theory.}
Clearly, in view of this equation, instability is not possible when $\widehat{f}_b-\overline{\rho}|\widehat{\kappa}_2|^2(\frac{s^2}{|U|^2+s^2}+\frac{a^2}{|U|^2})$ is positive everywhere, and this is satisfied if
\textcolor{black}{
\begin{eqnarray}
\widehat{f}_b-\overline{\rho}|\widehat{\kappa}_2|^2\left (1+\frac{a_M^2}{c_i^2} \right )\geq 0,\qquad a_M^2\equiv \max_{\Omega}a^2.\label{fbfb}
\end{eqnarray}
}

Let us meanwhile turn off the external magnetic effect ($a=0$) to focus on the effect of non-planar base flow configuration to the Miles-Howard stability condition. From (\ref{fbfb}), the stability of the flow is guaranteed if 
\begin{eqnarray}
\frac{\widehat{f}_b}{\overline{\rho}}-|\widehat{\kappa}_2|^2=[\phi^*, \psi^*]
\left [
\begin{array}{cc}
\overline{u}_y^2(J_1-\frac{1}{4}) & \overline{u}_y\overline{u}_z(J_{12}-\frac{1}{4}) \\
\overline{u}_y\overline{u}_z(J_{12}-\frac{1}{4}) & \overline{u}_z^2(J_2-\frac{1}{4})
\end{array}
\right ]
\left [
\begin{array}{c}
\phi\\ \psi
\end{array}
\right ]
~~~\label{phipsiq1}
\end{eqnarray}
is non-negative everywhere in $\Omega$.
Here we have defined the generalised Richardson numbers
\begin{eqnarray}
J_1\equiv \frac{N_1^2}{\overline{u}_y^2},\qquad J_2\equiv \frac{N_2^2}{\overline{u}_z^2},\qquad J_{12}\equiv \frac{N_{12}^2}{\overline{u}_y\overline{u}_z}.\label{JJJ}
\end{eqnarray}
The two eigenvalues $\lambda_+,\lambda_-$ of the matrix in (\ref{phipsiq1}) are found as
\begin{eqnarray}
\lambda_{\pm}=\frac{(\overline{u}_y^2+\overline{u}_z^2)(J-\frac{1}{4})\pm \sqrt{(\overline{u}_y^2+\overline{u}_z^2)^2(J-\frac{1}{4})^2-\overline{u}_y^2\overline{u}_z^2(2J_{12}-J_1-J_2)}}{2},~~~
\end{eqnarray}
where
\begin{eqnarray}
J\equiv \frac{N_1^2+N_2^2}{\overline{u}_y^2+\overline{u}_z^2}.\label{netJ}
\end{eqnarray}
Hence (\ref{phipsiq1}) is non-negative if $\lambda_+,\lambda_-\geq 0$, and those conditions are equivalent to
the Miles-Howard like condition 
\begin{eqnarray}
J_m\equiv \min_{\Omega}J \geq \frac{1}{4}\label{HM}
\end{eqnarray}
and
\textcolor{black}{
\begin{eqnarray}
J_{12} \geq \frac{J_1+J_2}{2},\label{JJ}
\end{eqnarray}
}
which is firstly noticed by
 Fung (1986). However, the question unanswered in that paper is that in what specific situation the condition (\ref{JJ}) is guaranteed.
 
Here we note that the identity $J_1J_2=J_{12}^2$ follows from (\ref{JJJ}).
This implies that $J_{12}$ is the geometric mean of $J_1$ and $J_2$, which must be smaller or equal to the arithmetic mean $(J_1+J_2)/2$. 
Therefore, clearly if 
(\ref{JJ}) is achieved only when the equal sign is established.
So what does the base flow look like when this condition is met?  
To answer this question, let us choose a point $(y,z)$ in the domain and consider the two local Cartesian coordinates attached to the contours of $\overline{u}$ and $G$. 
Let $n$ and $\nu$ be the coordinates normal to the contours of $\overline{u}$ and $G$, respectively. 
Noting that the contours of $\overline{\rho}$ and $G$ must coincide owing to (\ref{FUNGcond}), we have 
\begin{eqnarray}
2J_{12}-J_1-J_2=\frac{J}{\nu_y^2+\nu_z^2} \left \{2\left (\frac{\nu_y}{n_y}\right )\left (\frac{\nu_z}{n_z}\right )-\left (\frac{\nu_y}{n_y}\right )^2-\left (\frac{\nu_z}{n_z}\right )^2 \right\}, \label{J1212}
%
\end{eqnarray}
from the definitions of the Richardson numbers and the buoyancy frequencies.
Here the subscripts $n$ and $\nu$ represent the corresponding partial differentiations. The sum of the terms in the curly bracket cannot become positive because the eigenvalues of the matrix associated with the quadratic form are 0 and $-2$. 
This means that if the flow is stably stratified $2J_{12}-J_1-J_2\leq 0$, consistent to our aforementioned observation. 
We are interested in the case $2J_{12}-J_1-J_2=0$, because this is the only case (\ref{JJ}) is satisfied. When this happens the terms in the curly bracket in (\ref{J1212}) must vanish. This is possible only when $\frac{\nu_y}{n_y}=\frac{\nu_z}{n_z}$, namely the contours of $\overline{u}$ and $G$ are aligned (i.e. the gradients of $\overline{u}$ and $G$ are parallel and $J_1=J_2=J_{12}$). 
Therefore, the first important conclusion obtained in this section is that 
\textit{when $a=0$ the Miles-Howard condition (\ref{HM}) guarantees the stability of the flow only when the directions of the shear and the stratification are aligned everywhere in $\Omega$.} 

%
The second conclusion we shall derive here is that \textit{even if the above shear-stratification aligned condition is met, in the presence of an external magnetised field ($a\neq 0$) the Miles-Howard condition (\ref{HM}) no more guarantee the stability of the flow.}
Now let us assume the shear-stratification aligned condition $J_1=J_2=J_{12}$ for everywhere in $\Omega$.
We can write the left side of (\ref{fbfb}) in the quadratic form similar to (\ref{phipsiq1}),
and the argument analogues to the non-magnetised case leads $\lambda_+=(\overline{u}_y^2+\overline{u}_z^2)\{J-\frac{1}{4}(1+\frac{a_M^2}{c_i^2})\}$ and $\lambda_-=0$.
Recall that there is no unstable eigenvalue $c$ that makes $\lambda_+$ positive everywhere in $\Omega$. Hence $c_i^2(4J_m-1)\geq a_M^2$ must be satisfied for any unstable modes. Therefore, 
when $J_m>1/4$ we can derive the upper bound of $c_i$ 
\begin{eqnarray}
c_i^2 < \frac{a_M^2}{4J_m-1}.\label{HMstable}
\end{eqnarray}
Clearly, unless the external magnetic field is completely switched off, the stability of the flow cannot be guaranteed.


\subsection{Semi-ellipse type bound}

The outcome of section 4.1 can be combined with the semicircle theorem to yield a tighter bound. 
In this section we assume that the base flow is convectively not unstably stratified ($N_1^2+N_2^2\geq 0$ everywhere in $\Omega$). 

First we show that there is an alternative way to find the semicircle theorem.
Writing 
\begin{eqnarray}
\kappa=\eta_y+\zeta_z, \qquad \mathcal{D}=ik \xi+\kappa, \qquad \mathcal{G}=(G_y\eta+G_z\zeta), 
\end{eqnarray}
from (\ref{2q}) and (\ref{2xi}) we can deduce
\begin{eqnarray}
\overline{\rho}k^2 U^2 \xi
=-ik (s^2\overline{\rho}\mathcal{D}+\overline{\rho}\mathcal{G}),\label{xi4}
\end{eqnarray}
while (\ref{1eta}), (\ref{1zeta}) and (\ref{1rho}) become
\begin{eqnarray}
k^2\overline{\rho}(U^2-a^2) \eta=\widetilde{q}_y
+G_y(\eta \overline{\rho}_y+\zeta \overline{\rho}_z+\overline{\rho}\mathcal{D}),\label{eta4}\\
k^2\overline{\rho}(U^2-a^2) \zeta=\widetilde{q}_z
+G_z(\eta \overline{\rho}_y+\zeta \overline{\rho}_z+\overline{\rho}\mathcal{D}),\label{zeta4}\\
\widetilde{q}=-\overline{\rho}a^2\kappa-(s^2\overline{\rho}\mathcal{D}+\overline{\rho}\mathcal{G}).\label{qq4}
\end{eqnarray}
The integration by parts of $\eta^*\times$(\ref{eta4})$+\zeta^*\times$(\ref{zeta4}) over the domain becomes
\begin{eqnarray}
\langle k^2\overline{\rho}(U^2-a^2)(|\eta|^2+|\zeta|^2) \rangle =\langle \mathcal{G}^*(\eta \overline{\rho}_y+\zeta \overline{\rho}_z)+\overline{\rho}(\mathcal{G}^*\mathcal{D}+\kappa^*s^2\mathcal{D}+\kappa^*\mathcal{G}+a^2|\kappa|^2) \rangle,~~~~
\end{eqnarray}
after eliminating $\widetilde{q}$ using (\ref{qq4}).
Further using the identities $\mathcal{D}=\frac{U^2\kappa+\mathcal{G}}{U^2-s^2}$ and $\mathcal{G}^*\mathcal{D}+\kappa^*s^2\mathcal{D}+\kappa^*\mathcal{G}=\frac{s^2U^2|\kappa+s^{-2}\mathcal{G}|^2}{U^2-s^2}-\frac{|\mathcal{G}|^2}{s^2}$ led by (\ref{xi4}),
the integral above can be simplified as
\begin{eqnarray}
\langle U^2Q_2 \rangle =\left \langle \overline{\rho}\frac{s^2|U|^4|\kappa+s^{-2}\mathcal{G}|^2}{|U^2-s^2|^2}+f_b+\overline{\rho}a^2(k^2(|\eta|^2+|\zeta|^2)+|\kappa|^2)\right \rangle.\label{Q2eq}
\end{eqnarray}
Here 
\begin{eqnarray}
Q_2\equiv k^2\overline{\rho}(|\eta|^2+|\zeta|^2)+\overline{\rho}\frac{s^4|\kappa+s^{-2}\mathcal{G}|^2}{|U^2-s^2|^2}\label{Q2def}
\end{eqnarray}
 is a real and positive function, and
\begin{eqnarray}
f_b\equiv U^{-1/2}\widehat{f}_b
=
\overline{\rho}
[\eta^*, \zeta^*]
\left [
\begin{array}{cc}
N_1^2 & N_{12}^2 \\
N_{12}^2 & N_2^2
\end{array}
\right ]
\left [
\begin{array}{c}
\eta\\ \zeta
\end{array}
\right ]
\end{eqnarray}
is the buoyancy related term, which is non-negative for convectively stably stratified flows. 

The right side of (\ref{Q2eq}) is real and positive, and thus it is in a suitable form to derive a semicircle theorem. The argument similar to section 3.1 can be applied to (\ref{Q2eq}) to yield 
\begin{eqnarray}
\langle \{\overline{u}_-^2-(c_r-\overline{u}_+)^2-c_i^2-\Theta\}Q_2\rangle \geq \langle f_b\rangle,\label{semifb}
\end{eqnarray}
where a real number $\Theta$ is introduced to express the possible contraction of the semicircle radius from the usual case. 
From the inspection of the right side of (\ref{Q2eq}), the best choice of $\Theta$ can be found as
\begin{eqnarray}
\Theta=\min_{\Omega}\min \left (a^2,  \frac{c_i^4}{s^2}\right ).
\end{eqnarray}
(When there is no stratification at all ($G\equiv 0$) there is a better choice
$\Theta=\min_{\Omega} \min \left (a^2, c_T^2+\frac{a^2}{c_T^2}\frac{c_i^4}{s^2}\right )$
that can be found by using the identity $\frac{|U|^4}{s^2}+\frac{a^2}{s^4}|U^2-s^2|^2=c_T^2+\frac{s^2+a^2}{s^4}|U^2-c_T^2|^2$, but we do not go into further detail.) 

For some eigenvalues we can estimate $\langle f_b \rangle$ from below by a positive number, thereby further improving the bound. 
Relating $Q_1$ and $Q_2$ appropriately and using (\ref{compQ1}), after some laborious algebra given in Appendix E, we can find the estimate
\begin{eqnarray}
\langle f_b \rangle  \geq \left (J_m\frac{c_i^2}{\mathcal{C}^2}-\frac{\lambda_M}{k^2} \right )\langle Q_2\rangle, \label{estfb}
\end{eqnarray}
where
\begin{eqnarray}
\lambda_M\equiv 
\max_{\Omega}\frac{\sqrt{(\overline{u}_y^2+\overline{u}_z^2)^2(J-J_m)^2+4\overline{u}_y^2\overline{u}_z^2J_m(J_1+J_2-2J_{12})}-(\overline{u}_y^2+\overline{u}_z^2)(J-J_m) }{2},\label{lamM}~~~~~~\\
\mathcal{C}\equiv \frac{1+\sqrt{1+\mu(\frac{a_M^2}{c_i^2}-4J_m)}}{2\mu}.\label{defCCC}\hspace{40mm}
\end{eqnarray}
The estimate (\ref{estfb}) holds only when 
\begin{eqnarray}
\mu \equiv 1-\frac{\lambda_M}{k^2c_i^2}.\label{mulambda}
\end{eqnarray}
 is positive; otherwise we simply use the estimation $\langle f_b \rangle \geq 0$. 
The quantity $\lambda_M\geq 0$ measures how much the shear and the stratification are not aligned -- this is precisely what was missing in Fung (1986).
Combining (\ref{semifb}) and (\ref{estfb}), we have the bound
\begin{eqnarray}
(c_r-\overline{u}_+)^2+(\mu+\frac{J_m}{\mathcal{C}^2})c_i^2\leq \overline{u}_-^2 -\Theta.
\end{eqnarray}

We also remark that when $\mu>0$ the argument of the square root appeared in $\mathcal{C}$ should be positive (see (\ref{ABquad2})). 
From this condition, when $J_m>\frac{1}{4}$ we can deduce the following upper bound of $c_i^2$ (see Appendix F)
\begin{eqnarray}
c_i^2< \frac{(4J_m k^{-2}\lambda_M+a_M^2)+\sqrt{(4J_mk^{-2}\lambda_M+a_M^2)^2-4(4J_m-1)k^{-2}\lambda_M a_M^2}}{2(4J_m-1)}.\label{upperci}
\end{eqnarray}

The above results can be summarised in the following theorem.
\begin{thm}
If $N_1^2+N_2^2 \geq 0$ everywhere in $\Omega$, the unstable complex phase speed $c=c_r+ic_i$ of (\ref{full_linear}) is bounded so that 
\begin{eqnarray*}
 (c_r-\overline{u}_+)^2+\varGamma c_i^2 \leq \overline{u}_-^2-\min_{\Omega}\min \left (a^2,\frac{c_i^4}{s^2}\right )
\end{eqnarray*}
with
\begin{eqnarray*}
\varGamma=
\left \{
\begin{array}{c}
\max\left (1,\mu+4\mu^2J_m \left \{1+\sqrt{1+\mu\left (\frac{\max_{\Omega}a^2}{c_i^2}-4J_m\right )}\right \}^{-2}\right )
~~~\text{if}~~~\mu>0,\\
1~~~~\text{otherwise},
\end{array}
\right .
\end{eqnarray*}
where $J_m=\min_{\Omega}J$ and $\mu$ is the quantity defined in (\ref{mulambda}).
%
%
If $J_m>\frac{1}{4}$, the inequality (\ref{upperci}) must also be satisfied.
%
\end{thm}

 
 
 
 \begin{figure}
\centering
\includegraphics[scale=1.]{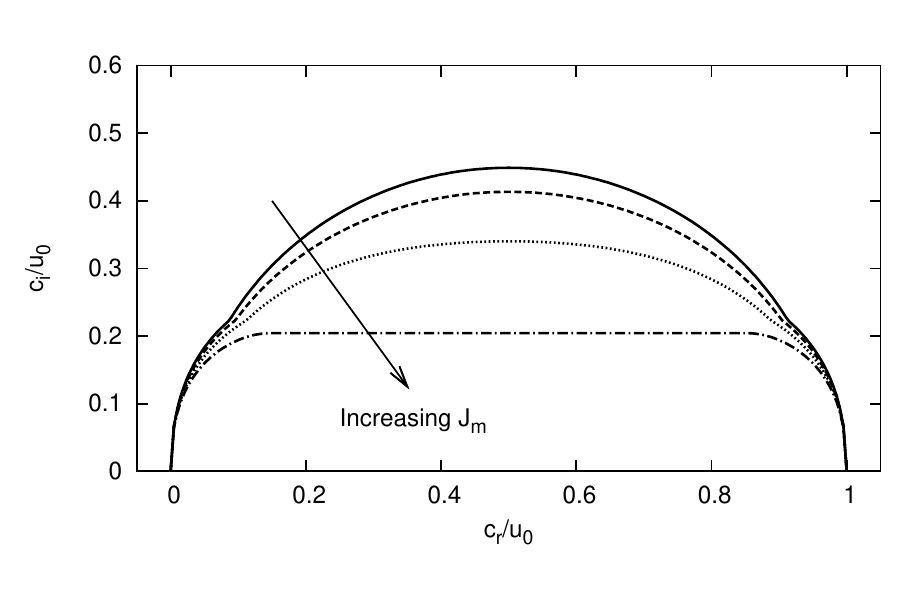}
\caption{The eigenvalue bound (\ref{eboundsimple}). The parameters used are
$\overline{u}_+=\overline{u}_-=u_0/2$, $ a^2=0.025u_0^2$ and $s^2=0.1u_0^2$. 
The solid, dashed, dotted and dot dashed curves correspond to $J_m=0.1,0.2,0.3,0.4$,  respectively. The usual semicircle bound is the circle of diameter unity centred at $(0.5,0)$.
}
\label{fig:longwave}
\end{figure}

The expression of the bound is rather complicated, so it might be worth considering the perfectly shear-stratification aligned case first (namely $\lambda_M=0$).
For the sake of simplicity, we further assume that the Alfv\'en and sound speeds are constants.
In this case, the bound in theorem 2 can be simplified as
\begin{subequations}\label{eboundsimple}
\begin{eqnarray}
 (c_r-\overline{u}_+)^2+c_i^2\left \{\frac{4J_m}{\left (1+\sqrt{1+\frac{a^2}{c_i^2}-4J_m}\right )^2}+1\right \} \leq \overline{u}_-^2-\min \left (a^2,  \frac{c_i^4}{s^2}\right ),\label{eboundsimple1}\\
\text{additionally}~~~~
c_i^2< \frac{a^2}{4J_m-1}\qquad \text{if} \qquad J_m>\frac{1}{4}.\label{eboundsimple2}
\end{eqnarray}
\end{subequations}
In order to plot the bound in the complex plane, we set $(a^2,s^2)=(0.025u_0^2,0.1u_0^2)$, where $u_0$ is a constant specifying the range of the base velocity ($\overline{u} \in [0,u_0]$).
The eigenvalue bounds (\ref{eboundsimple}) for $J_m=0.1,0.2,0.3,0.4$ are shown in figure 3. We can see that the boundary becomes tighter as the effect of the stratification gets stronger. 
The bounds for $J_m=0.1,0.2$ are determined solely by (\ref{eboundsimple1}) because $J_m\leq \frac{1}{4}$. The dents on the bounds are due to the two possible values of the right side of (\ref{eboundsimple1}), $\overline{u}_-^2-a^2$ and $\overline{u}_-^2-c_i^4/s^2$; the middle part of the bound is determined by the former and the near ends by the latter.
While when $J_m>\frac{1}{4}$ we must also consider (\ref{eboundsimple2}), which is essentially (\ref{upperci}) for the general case.
(Note also that this condition reduces to (\ref{HMstable}) for the perfectly shear-stratification aligned case.)
For $J_m=0.3$ the upper bound of $c_i/u_0$ predicted by (\ref{eboundsimple2}) is about $0.355$, which is well above the bound determined by (\ref{eboundsimple1}). Thus for this case, the overall picture of the bound is unchanged from the previous two cases.
However, for $J_m=0.4$, the upper bound (\ref{eboundsimple2}) does play a role in the eigenvalue bound. 
In the figure, the flat top of the bound at $c_i/u_0\approx 0.205$ corresponds to this upper bound. 
\begin{figure}
\hspace{-35mm}(a)\\
\centering
(b)\includegraphics[scale=1.]{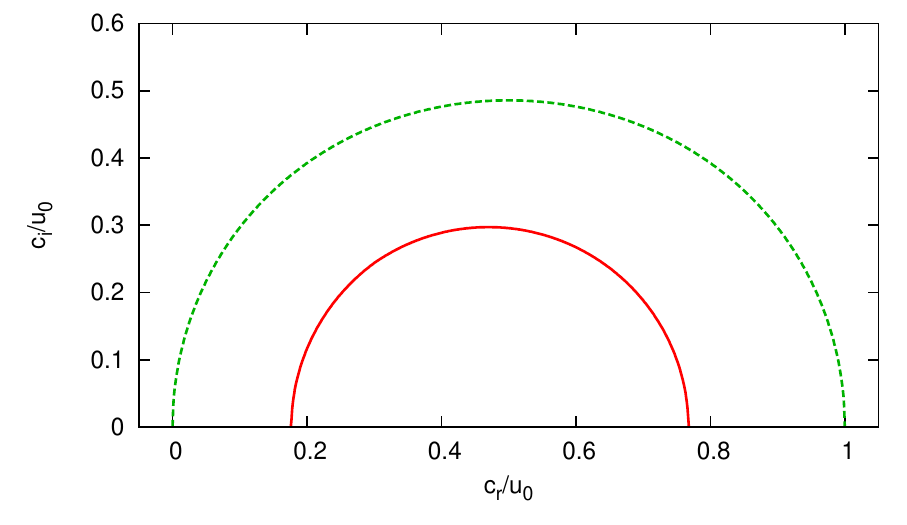} \\
\includegraphics[scale=1.]{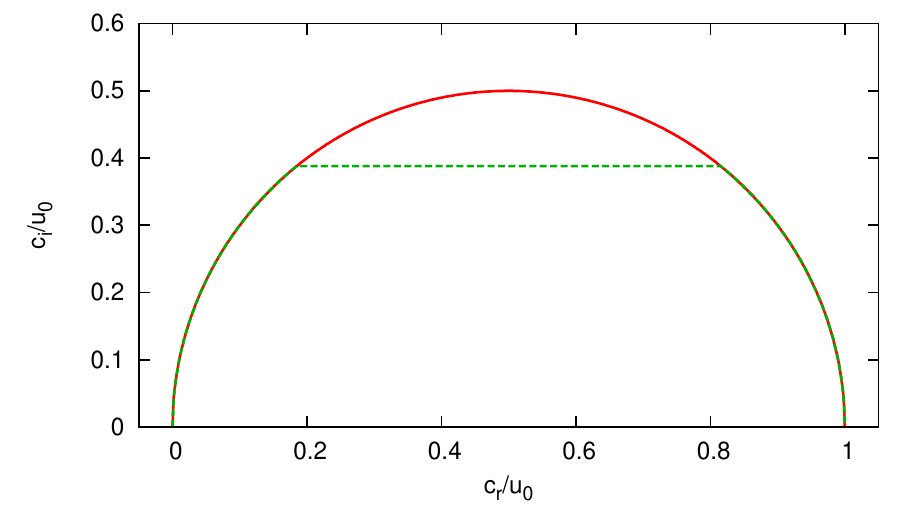}
\caption{The green dashed curves are the semi-ellipse eigenvalue bounds by theorem 2 for the model flow defined in section 2.4. The red solid curves are the inner envelope bounds (theorem 1). Note that both bounds should always fall within the usual semicircle bound (the circle of diameter unity centred at $(0.5,0)$).
(a): $(k,u_0,g,\beta_0)=(10, 0.5, 1, 30)$. The semi-ellipse bound is almost the usual semicircle bound because $J_m$ is small ($J_m=0.0118$ and $\lambda_M=0.471$). (b): $(k,u_0,g,\beta_0)=(40, 0.5,10,10^6)$. $J_m=1.18$ and $\lambda_M=47.1$. 
}
\label{fig:longwave}
\end{figure}




The base flow defined in section 2.4 is a perfect model to study the effect of non-zero $\lambda_M$, because as seen in figure 1 the directions of the shear and the stratification are not aligned. Figure 4 is the comparison of the bounds determined by theorem 1 (red solid) and theorem 2 (green dashed). 
Figure 4a is the result for the parameters used in figure 2a, $(k,g,\beta_0)=(10,1,30)$. In this case, the inner envelope bound gives a tighter bound than the semi-ellipse type bound. This is an expected result because the minimum Richardson number of this flow is small ($J_m=0.0118$), and so there is not much stabilisation effect by the stratification. 
In figure 4b, we changed the parameters to $(k,g,\beta_0)=(40,10,10^6)$. 
The semi-ellipse bound becomes more efficient because the value of $g$ is larger than the previous case, and hence we have a stronger stable stratification. 
The minimum Richardson number is $J_m=1.18>\frac{1}{4}$ so the upper bound defined by (\ref{upperci}) appears as the flat top of the boundary.
Of course, this upper bound must be small enough to be effective. 
As seen in the simplified case (\ref{eboundsimple2}) it is desirable to choose a small magnetic field for this purpose, as we did in figure 4b.
Also, in order to see the effect of the upper bound of $c_i$ in the eigenvalue bound, $\lambda_M/k^2$ must be sufficiently small. 
To see this, recall that in order for the stabilising effect of stratification to appear $\mu=1-\lambda_M/k^2c_i^2$ must be positive.
Recall that the semi-ellipse bound is always tighter than the usual semicircle bound, and thus
we have the estimation $c_i^2\leq \overline{u}_-^2$.
This means that at least the inequality $\lambda_M <k^2\overline{u}_-^2$ must hold to see the the effect of (\ref{upperci}); for example in figure 4a $\lambda_M=47.1$ and $k^2\overline{u}_-^2=100$. 
An important feature of the non-planar cases is that the stabilising effect of stratification is lost when the wavenumber is small.







\subsection{Outer envelope type bound}

\begin{figure}
\centering
\includegraphics[scale=1.]{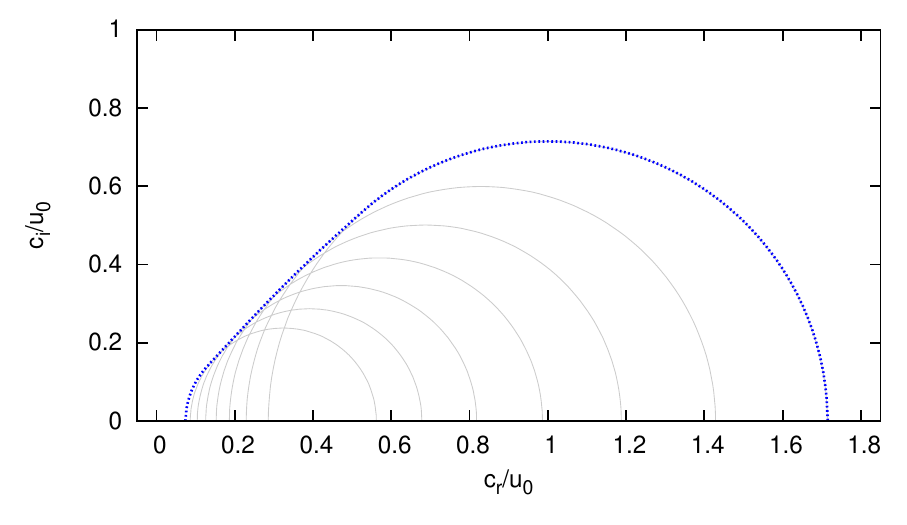}
\caption{The blue dotted curve is the eigenvalue bound by theorem 3 for the model flow defined in section 2.4. 
This is the outer envelope of the grey thin curves, which correspond to the semicircle type inequalities for various fixed $(y,z)$.
The choice of the parameters are $(k,u_0,g,\beta_0)=(10, 0.5, 1, 30)$. The usual semicircle bound is the circle of diameter unity centred at $(0.5,0)$.
}
\label{fig:longwave}
\end{figure}
From the examination of the integral (\ref{compQ1}), clearly
\begin{eqnarray}
\left (1+\frac{a^2}{|U|^2} \right )k^2 (|\phi|^2+|\psi|^2)+\frac{\widehat{f}_b}{\overline{\rho}|U|^2}-\frac{|\widehat{\kappa}_2|^2}{|U|^2}\left (\frac{s^2}{|U|^2+s^2}+\frac{a^2}{|U|^2} \right )<0\label{appc1}
\end{eqnarray}
must be satisfied somewhere in the flow to generate an unstable mode ($\widehat{f}_b$ and $\widehat{\kappa}_2$ are defined in (\ref{defkappa}), (\ref{deffb})). 
This inequality for the function of $(\phi,\psi,|U|^2)$ is the key to find alternative proof of the eigenvalue bound derived by Cally (1983). This new method is motivated by Howard (1961), who found a growth rate bound as a by-product of the derivation of the Miles-Howard condition. The good thing about this alternative approach is that it can be extended to the general non-planar cases as well. 

Let us fix $(y,z) \in \Omega$.
The left hand side of (\ref{appc1}) can be written in the quadratic form
\begin{eqnarray}
[\phi^*, \psi^*]
\left [
\begin{array}{cc}
A+N_1^2-\chi \overline{u}_y^2 & N_{12}^2-\chi \overline{u}_y\overline{u}_z \\
N_{12}^2-\chi \overline{u}_y\overline{u}_z  & A+N_2^2-\chi \overline{u}_z^2
\end{array}
\right ]
\left [
\begin{array}{c}
\phi\\ \psi
\end{array}
\right ],
%
\end{eqnarray}
using
\begin{eqnarray}
A(|U|^2)=k^2(|U|^2+a^2),\qquad \chi(|U|^2)=\frac{(s^2+a^2)(|U|^2+c_T^2)}{4|U|^2(|U|^2+s^2)}.
\end{eqnarray}
The eigenvalues of the above matrix are
\begin{eqnarray}
F_{\pm}(|U|^2)=A+\frac{1}{2}(B\pm \sqrt{B^2+4C}),\label{Fpm}
\end{eqnarray}
where 
\begin{eqnarray}
B(|U|^2)=(N_1^2+N_2^2)-\chi (\overline{u}_y^2+\overline{u}_z^2),\\
C(|U|^2)=\chi (\overline{u}_y^2N_2^2-2\overline{u}_y\overline{u}_zN_{12}^2+\overline{u}_z^2N_1^2).
\end{eqnarray}
From this result we see that the inequality (\ref{appc1}) can be written in the form $F_+\times (\text{positive quantity})+ F_-\times (\text{positive quantity})<0$. Since $F_- \leq F_+$, in order to have an unstable mode the inequality $F_-(|U|^2)<0$ must be satisfied at least one point $(y,z)$ in $\Omega$. 


For sufficiently large $|U|^2$, the function $F_-(|U|^2)$ seen in (\ref{Fpm}) behaves like $k^2|U|^2$ and hence it monotonically increases. 
Now let us denote the largest root of that function as $R^2$ (i.e. $F_-(R^2)=0$). For $|U|^2\geq R^2$ the condition $F_-(|U|^2)<0$ cannot be satisfied.
Therefore, for any unstable modes the semicircle type inequality
$(\overline{u}-c_r)^2+c_i^2=|U|^2<R^2$
must be satisfied at least one point $y,z$ in the domain. 
\begin{thm}
The unstable complex phase speed $c=c_r+ic_i$ of (\ref{full_linear}) must satisfy
\begin{eqnarray}
(\overline{u}(y,z)-c_r)^2+c_i^2\leq \{R(y,z)\}^2\nonumber
\end{eqnarray}
for some $(y,z)\in \Omega$. Here $R^2$ is the largest positive root of $F_-(R^2)=0$, where the function $F_-$ is defined in (\ref{Fpm}).
\end{thm}
This theorem says that the eigenvalues must lie within at least one of the semicircles $(\overline{u}(y,z)-c_r)^2+c_i^2= \{R(y,z)\}^2$ calculated for various fixed $(y,z)$.
Therefore, in practice, the net eigenvalue bound can be found by plotting the semicircles in the complex plane and taking the \textit{outer envelope} of them. 
(Note the difference to theorem 1, where the eigenvalues must lie within \textit{all} the semicircles.)
Figure 3 demonstrates how theorem 3 can be used to construct a net eigenvalue bound for the model flow defined in section 2.4.
The parameters in the flow are the same as those used in figures 2a and 4a.
The overall property of the outer envelope bound is quite different from the previous two bounds based on the Howard semicircle theory, and thus by using them in a complementary manner, a tighter net bound could be obtained. We will clarify the basic property of the outer envelope bound in the next section.

It is straightforward to check that the above result is in fact the generalisation of the Cally (1983) theory.
When the base field is independent of $y$, of course $C=0$. In this case the inequality $F_-(|U|^2)<0$ reduces to (note that we are only interested in the case $B< 0$ where $F_-$ may become negative)
\begin{eqnarray}
k^2(|U|^2+a^2)+N_2^2-\frac{(s^2+a^2)(|U|^2+c_T^2)}{4|U|^2(|U|^2+s^2)}u_z^2<0,
\end{eqnarray}
which is the inequality (3.19) derived in Cally (1983).
In that paper the perturbations are limited to those that are independent of $y$ to derive an eigenvalue bound in the complex plane.
However, clearly this limitation can be removed from the above result.

As noted by Cally (1983), for non-magnetised cases, \textcolor{black}{the bound reduces to} the stability condition by Chimonas (1970), who extended the Miles-Howard stability condition for compressible flows. The same argument does not hold for the non-planar flows due to the existence of the term $C$, and this is consistent with what we obtained in section 4.1.

\section{Comparison of the bounds and numerical eigenvalues}

Here we shall compare the eigenvalue bounds obtained so far and the numerical eigenvalues of (\ref{full_linear}). In order to facilitate the calculation of the eigenvalues of the system, we use simple planar model flow configurations. 
The advantage of using the planar problems is that the stability problem becomes ordinary differential equations and thus the numerical investigation is not difficult. On the other hand, in the general non-planar case, partial differential equations must be used, which makes the calculations much more challenging (see section 2.3).
Another motivation for treating the planar problems is to show that the results of this paper contain new findings even for the classical cases.






We consider three base flow configurations in a layer $z\in[0,1]$ subjected to a downward gravity field $G=-gz$, as summarised in figure 6. 
The choice of the adiabatic exponent $\gamma$ is 5/3.
Since the base flows are not depending on $y$, the $y$ dependence of the perturbation can be written by a monochromatic Fourier mode without loss of generality. 
Here the perturbation is assumed to be proportional to $\exp (ikx+ily-ikct)$ with the spanwise wavenumber $l$. The shooting method is used to compute the eigenvalue $c$ of (\ref{full_linear}).


\begin{figure}
(a) \hspace{60mm} (b)\\
\includegraphics[scale=1.]{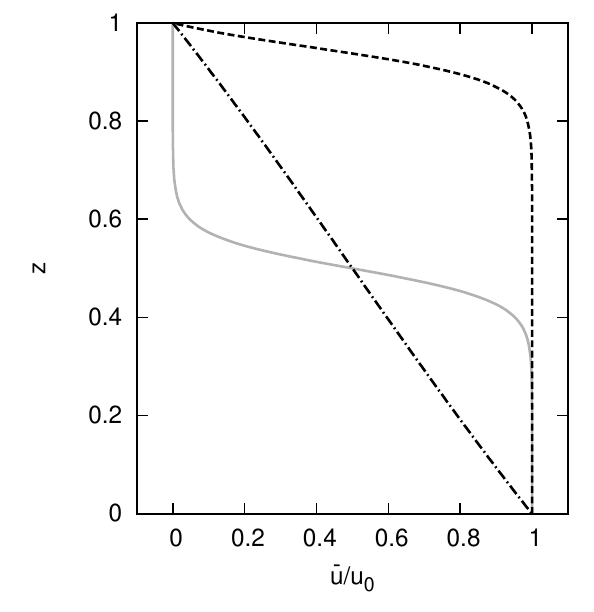} 
\includegraphics[scale=1.]{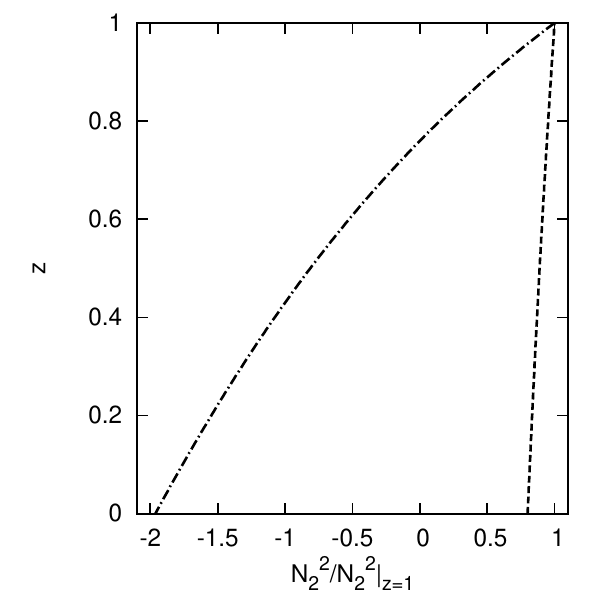}
 \caption{The base flows used in the comparison. The grey solid curve in (a) shows the hyperbolic tangent profile (\ref{tanh}) with $\alpha=15, z_s=0.5$. This profile is relevant for figure 7, where the polytrope atmosphere with $m=3/2$ is used (thus $N_2^2=0$). The dashed curve is the same base flow but $m=2, z_s=0.95$, which is used in figure 8. The corresponding buoyancy frequency profile is given in (b).
The dot-dashed lines in (a,b) are the base profiles used in figure 5 of Tobias \& Hughes (2004) and figure 9 of this study.}
\end{figure}

Let us begin our analysis by introducing the well-known hyperbolic tangent shear profile 
\begin{eqnarray}
\overline{u}=u_0\frac{\tanh(\alpha(1-z_s))-\tanh(\alpha(z-z_s))}{\tanh(\alpha(1-z_s))+\tanh(\alpha z_s)}. \label{tanh}
\end{eqnarray}
This profile produces an inviscid hydrodynamic instability of the Rayleigh (or Kelvin-Helmholtz) type due to the existence of the inflection point at $z=z_s$ in the flow. The shear there can be controlled by the parameter $\alpha$, while the velocity range is $[0,u_0]$.
The shear profile (\ref{tanh}) is considered in a polytropic atmosphere subjected to a uniform external magnetic field. Such an atmosphere has widely been used in various astrophysical magneto-atmospheric problems; see Bogdan \& Cally (1997) and references therein.
Here we set $\overline{\rho}^{1+1/m}=\gamma \overline{p}$ with the polytrope index $m$, choosing the proportional constant so that the sound wave speed $s$ at $z=0$ becomes unity.
The magneto-static condition $\frac{d\overline{p}}{dz}+g\overline{\rho}=0$ with $\overline{\rho}(0)=1$ yields
\begin{eqnarray}
\overline{\rho}=(1-\frac{z}{L})^m,\qquad \overline{p}=\frac{1}{\gamma}(1-\frac{z}{L})^{m+1}.
\end{eqnarray}
Here $L=\frac{m+1}{\gamma g}$ and we fix the constant $g$ so that $L=5$. 
The squares of the Alfv\'en and sound wave speeds are found as
\begin{eqnarray}
a^2=\frac{2}{\gamma \beta_0} \frac{1}{(1-\frac{z}{L})^m}, \qquad s^2=(1-\frac{z}{L}),
\end{eqnarray}
where $\beta_0$ is the plasma beta at $z=0$. Using (\ref{N12}), $N_1^2=0$ and
\begin{eqnarray}
N_2^2=\frac{g}{L-z}(m-\frac{m+1}{\gamma}).
%
\end{eqnarray}
The control parameters of the flow are $m, u_0,\alpha, z_s,$ and $ \beta_0$.

First let us consider the parameters $(m, u_0,\alpha, z_s, \beta_0)=(3/2,0.5, 15, 0.5, 150)$, setting the rigid boundary conditions at $z=0$ and 1. 
The index $m=3/2$ is the special case where the buoyancy frequency $N_2^2$ becomes zero everywhere in the flow. The plasma beta of the flow is large, meaning that the imposed magnetic field is weak. Thus we expect that the situation is not too far from the usual hydrodynamic unstratified shear flow problem. The inflection point is placed at the middle of the flow domain.
The growth rate takes its maximum around $k=10, l=0$, and for this case, we compared it with the bounds in figure 7a. 
The complex phase speed is normalised by the velocity range $u_0$ in the figure so that the usual semicircle bound becomes the circle of diameter unity centred at $(0.5,0)$. 
The phase speed $c_r$ of the numerical eigenvalue (magenta cross) is about half of $u_0$, and with respect to the $c_r/u_0=0.5$ axis, the plot is almost symmetric.
As anticipated, the inner envelope bound (theorem 1, the red solid curve) and the semi-ellipse bound (theorem 2, the green dashed curve) are very close to the usual semicircle bound. 
The blue dotted curve is the outer envelope bound (theorem 3), which now reduces to the bound obtained by Cally (1983). 
This bound gives a tighter growth rate estimate than the other two bounds in figure 7a, but it typically becomes looser when the shear is strong, or the wavenumber $k$ is small (see figure 7b,c). 
This property is very similar to the eigenvalue bound (4.1) derived in Howard (1961); in fact, Cally's bound can be reduced to this bound in the absence of the compressibility and the magnetohydrodynamic effect. 

\begin{figure}
\hspace{5mm} (a) \hspace{35mm} (b)\hspace{35mm} (c)\\
\includegraphics[scale=0.8]{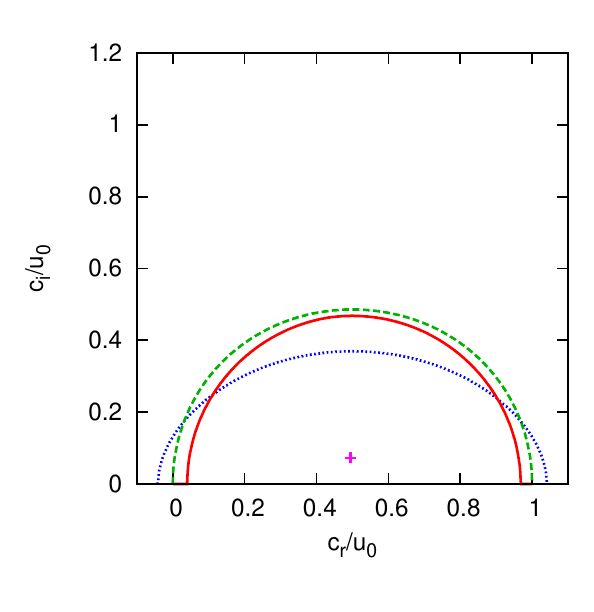} \hspace{-10mm}
\includegraphics[scale=0.8]{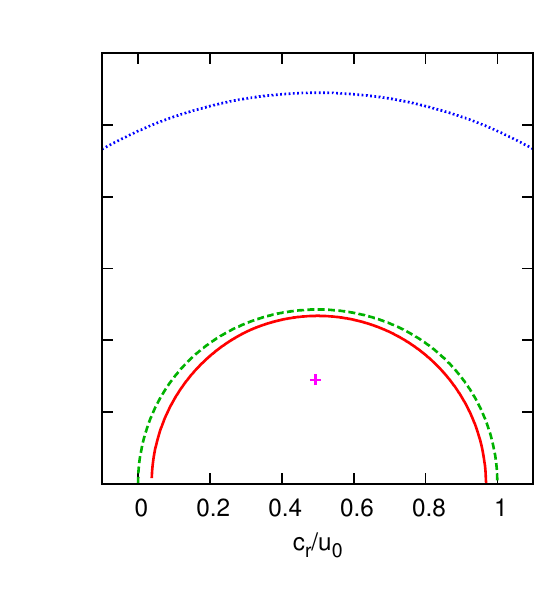} \hspace{-10mm}
\includegraphics[scale=0.8]{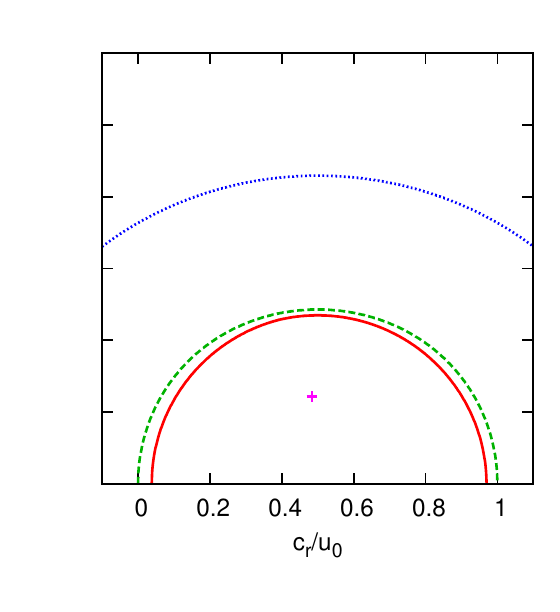}
 \caption{The comparison of the eigenvalues (magenta crosses) and the eigenvalue bounds for the hyperbolic tangent shear profile (\ref{tanh}) with $(u_0,\alpha, z_s, \beta_0)=(0.5, 15, 0.5, 150)$. 
 The convectively neutrally stable polytrope atmosphere ($m=3/2$) is used. The external magnetic field is uniform.
 The red solid, the green dashed, and the blue dotted curves are the bounds obtained in theorems 1, 2, and 3, respectively.
(a) is the result for $\alpha=15, k=10,l=0$.
The other panels are the same result but in (b) the larger shear parameter $\alpha=50$ is used, while in (c) the smaller wavenumber $k=4$ is used. 
 Note that the usual semicircle bound is the circle of diameter unity centred at $(0.5,0)$.}
\end{figure}
Next, we introduce some more asymmetry in the flow, changing the inflection point $z_s$ to $0.95$, and the boundary condition at the top ($z=1$) to the free one.
Also, we select $m=2$, so that the buoyancy frequency varies in $z$. As shown in figure 6b, the flow is convectively stably stratified. 
The maximum growth rate can still be found at $l=0$ as typical for shear dominated instabilities (however, note that Squire's theorem is formally not valid for magnetised problems; see Hunt (1966)).
The comparison is given in figure 8, where the definitions of the curves and points remain the same as those in figure 7. 
The magenta crosses in Figure 8a show how the value of $c_i$ changes in the numerical eigenvalue problem when $u_0$ is varied; the result is compared with 
the maximum $c_i$ predicted by the bounds. 
Figures b,c are the comparison of the numerical eigenvalues and the bounds in the complex plane at $u_0=0.6$ and $1.6$, respectively. 
In figures b,c, what is immediately apparent is that the inner envelope and Cally's bounds become largely asymmetric -- the property that was not seen in the usual semicircle bound. As the flow is convectively stably stratified, the inner envelope bound becomes tighter than the usual semicircle bound (see section 3.2). 
The contraction of the bound effectively works because the wavenumber is larger than that associated with the density scale height ($k_h \in [0.36, 0.45]$).
The semi-ellipse bound is also tighter than the usual semicircle bound but looser than the inner envelope bound, because the stabilisation effect by the stratification is not apparent due to the strong imposed magnetic field ($\beta_0=50$).
Turning back to Figure 5a, we can see that for $u_0\gtrsim0.87$ Cally's bound gives a tighter bound in terms of $c_i$. While for small $u_0$ the inner envelope bound gives a better result, predicting the critical value $u_0\approx 0.36$ below which the flow must be stabilised. 
This critical value is not too far from the numerically found stability threshold $u_0\approx 0.58$ at which the unstable eigenvalue vanishes.

\begin{figure}
(a)\\
\centering
\includegraphics[scale=1.]{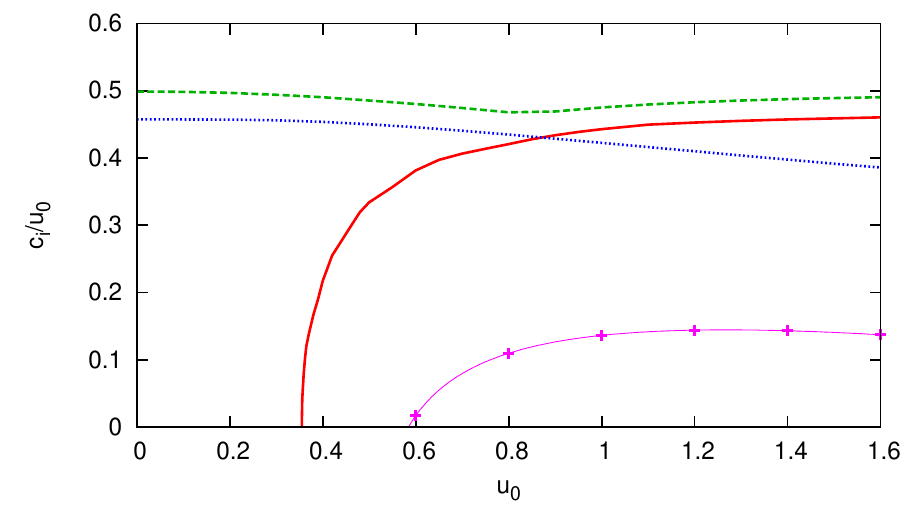}  \\
\hspace{-40mm}(b)\hspace{50mm} (c) \hspace{20mm} \\ \vspace{-10mm}
 \begin{tabular}
[c]{cc}
\includegraphics[scale=1.]{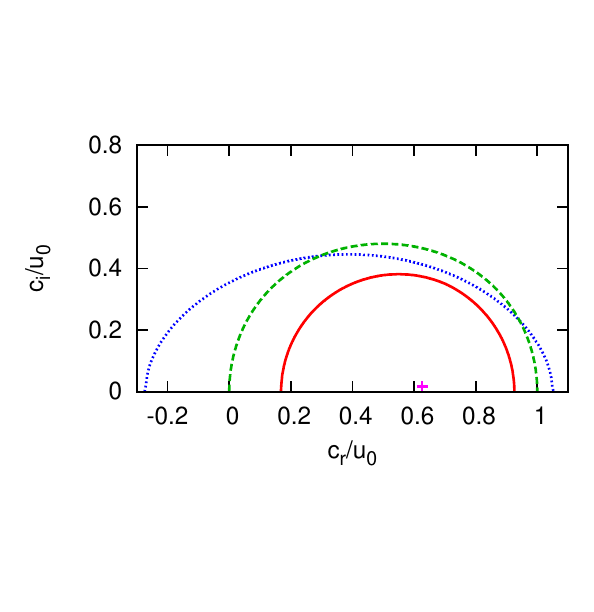} &
\includegraphics[scale=1.]{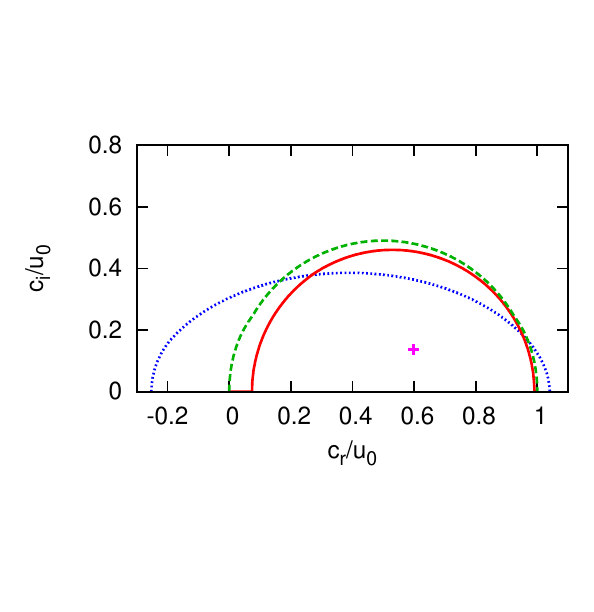} 
\end{tabular}
\vspace{-10mm}
 \caption{
 The eigenvalues and the eigenvalue bounds for the hyperbolic tangent shear profile (\ref{tanh}) with $(k,l,\alpha, z_s, \beta_0)=(10,0,15, 0.95, 50)$.
  The convectively stable polytrope atmosphere with $m=2$ is used. The external magnetic field is uniform.
The definitions of the curves and the points are the same as figure 7.
 (a): the comparison of the numerical eigenvalues (magenta crosses) and the upper bounds of $c_i$ predicted by theorems 1,2 and 3. (b) and (c) are the comparison in the complex plane at $u_0=0.6$ and $1.6$, respectively.
 %
}
\end{figure}

\begin{figure}
(a)\\
\centering
\includegraphics[scale=1.]{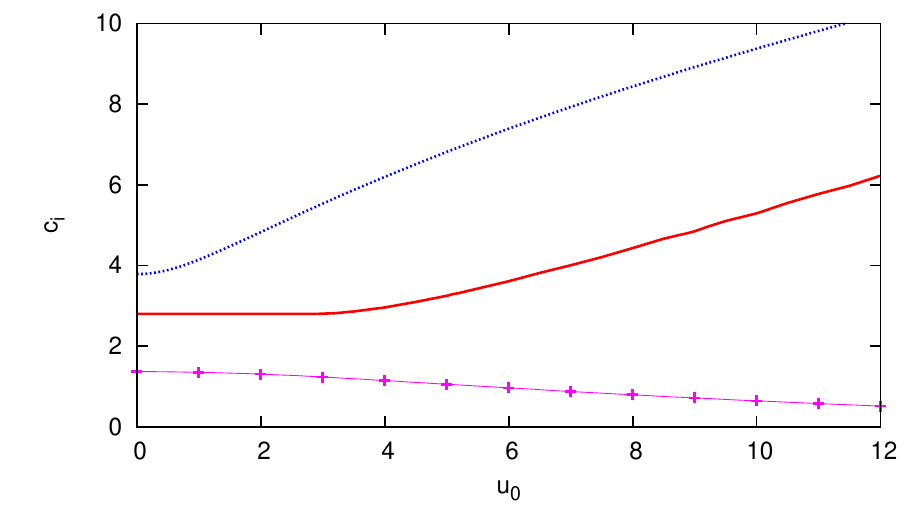}   \\
\hspace{-40mm}(b)\hspace{50mm} (c) \hspace{20mm} \\ \vspace{-10mm}
 \begin{tabular}
[c]{cc}
\includegraphics[scale=1.]{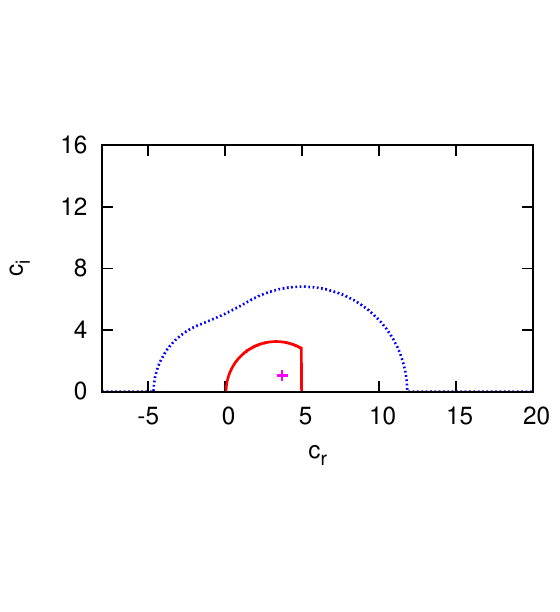} &
\includegraphics[scale=1.]{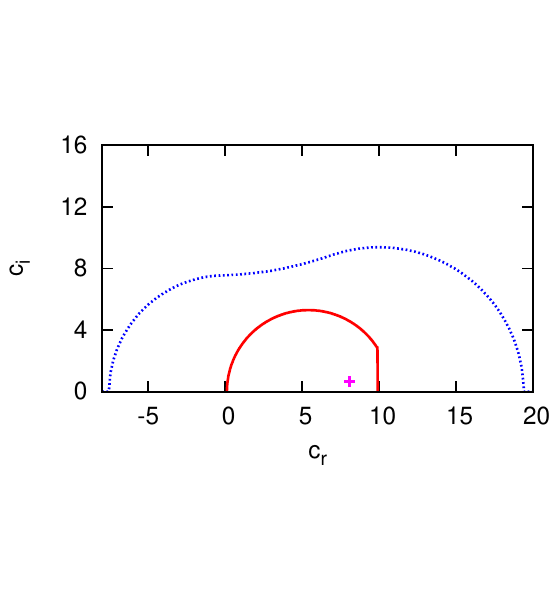} 
\end{tabular}
\vspace{-10mm}
 \caption{
(a): The same comparison of $c_i$ as figure 8a but for the flow studied in figure 5 of Tobias \& Hughes (2004). 
The shear profile (\ref{cubicshear}) is used with $(k,l,g,\beta_0)=(0.32,8,10,5)$.
The atmosphere is isothermal, and the linearly varying magnetic field generates the magnetic buoyancy instability.
 (b) and (c) are the comparison in the complex plane at $u_0=5$ and $10$, respectively (see figure 7 for the definitions of the points and the curves).
}
\end{figure}

In the third flow configuration, we study the convectively unstable case. For this purpose, we simply employ the model flow configuration studied in Tobias \& Hughes (2004) considering the linearly varying magnetic field proportional to $1+2.8 (1-z)$ and the isothermal atmosphere $\overline{p}=\frac{\beta_0}{2}\overline{\rho}$. Here the constant $\beta_0$ is chosen so that it is the plasma beta of the flow at $z=1$.
The density profile can be found by
the magneto-static condition
$0=\frac{\beta_0}{2}\frac{d\overline{\rho}}{dz}+g \overline{\rho} +\frac{1}{2\mu_0}\frac{d\overline{B}^2}{dz}$
together with the boundary condition $\overline{\rho}(1)=1$.
For $z\gtrsim 0.76$ the flow becomes convectively unstable because the buoyancy frequency $N_2^2$ becomes negative; see figure 6b.
Since this phenomenon is caused by the application of the non-uniform magnetic field, it is called magnetic buoyancy instability. The squares of the Alfv\'en and sound wave speeds are found as
\begin{eqnarray}
a^2=\frac{\{1+2.8 (1-z)\}^2}{\overline{\rho}},\qquad s^2=\frac{\gamma \beta_0}{2}.
\end{eqnarray}
We employ the shear profile 
\begin{eqnarray}
\overline{u}=u_0 \left (\frac{(1-z)^3}{6}-\frac{(1-z)^2}{4}+\frac{13(1-z)}{12} \right )
\label{cubicshear}
\end{eqnarray}
that was used in figures 4 and 5 of Tobias \& Hughes (2004). This profile is almost linear, but possesses an inflection point at $z=0.5$; see figure 6a.
The case $u_0=0$ corresponds to the flow studied in Hughes \& Cattaneo (1987). 
(In those papers $z$ is pointing downwardly so that $z=0 $ and $1$ are the upper and lower boundaries, respectively. Thus the shear and magnetic field profiles used there look a bit different to those given above. 
Note also that the parameters $(\delta, \lambda, \tilde{\beta})$ in Tobias \& Hughes (2004) correspond to our $(u_0, g, \beta_0/2)$. )

The parameters $(k,l,g,\beta_0)=(0.32,8,10,5)$ are selected in figure 9a so that we have the same flow configuration studied in figure 5 of Tobias \& Hughes (2004). 
Figures 9b,c are the plot in the complex plane at  $u_0=5,10$. 
One of the important differences from the previous cases is that the spanwise wavenumber $l$ is not zero, and hence the flow configuration is outside the scope of the semicircle theorem by Adam (1978b) and the eigenvalue bound by Cally (1983). 
A natural question here is that whether the usual semicircle theorem deduced in section 3.1, extending  Adam's result, can be used or not.
The answer is no because the quantity $k^2a^2+N_2^2$ is not always positive in the flow (see the discussion just below (\ref{NNNN})).
The fact that the theorem actually does not hold can be confirmed in the numerical result shown in figure 6a because 
the numerically obtained $c_i$ does not tend to zero as $u_0\rightarrow 0$ (this is the reason why we did not use the normalisation by $u_0$ in this figure).
The semi-ellipse theorem (theorem 2) cannot be used for this model flow as well, because the flow is convectively unstable.
Therefore, the only eiganvalue bounds that are available in figure 6 are the inner envelope bound (theorem 1) and the outer envelope bound (theorem 3).
Those bounds are depicted by the green dashed and red solid curves, respectively, in the figures.
Both of them are consistent with the numerical eigenvalues, and the former gives tighter results than the latter.


\section{Discussion and conclusions}

We have established three new eigenvalue bounds for the ideal instability of compressible stratified magneto-shear flows varying in two transverse directions. 
The two bounds obtained are the improved/generalised versions of the Howard semicircle theorem, and the other bound is similar to that derived by Cally (1983) for the planar cases. The properties of the bounds were studied by using the model flow through a duct of non-circular cross-section (figure 1). 

The first bound, called the inner envelope bound (theorem 1), is based on the energy principle of the Lagrangian displacement and treats the centre of the semicircle as a parameter. By considering the change in the potential energy owing to the Galilean transformations, we can derive the radii of the semicircles $R(r_c)$ for a given centre $r_c$ on the real axis. 
Since the unstable eigenvalues should lie in the semicircles for any $r_c$ in the complex plane, we can grasp the whole picture of the net eigenvalue bound by drawing an inner envelope of them. 
Unlike the usual semicircle theorem, the inner envelope theory does not require any particular conditions, and hence can be applied even for convectively unstable flows.
In the case of convectively stable stratification, the inner envelope theory should give a bound tighter than the semicircle bound, with a novel property that for some cases the stability of the flow can be established.
This stabilisation is essentially the well-known effect by the applied magnetic field, but 
a notable finding is that for the compressible stratified flows the stabilisation is effective only when the wavelength is sufficiently smaller than the density scale height.



The main idea used in the first type bound is so simple that it can be applicable for many other flows. Of particular interest is the extension of the theory to spherical coordinates because such extended theory may have some relevance to the study of solar/planetary atmospheres.
For two-dimensional purely hydrodynamic flows, some extensions of the usual semicircle theorem to the spherical coordinates have been attempted in the geophysics community; see Thuburn \& Haynes (1996), Sasaki et al. (2012).
Another interesting problem would be 
the stability analysis of viscoelastic flows, where it has been pointed out that there is an analogy to 
the magnetohydrodynamic problems (Ogilvie \& Proctor (2003)).

The second and third types of bounds are deduced by considering the generalisation of the Miles-Howard stability condition. The flow considered in this paper is subjected to a general conservative force field varying in two directions, as did in Fung (1986).
While such a generalisation might seem to complicate the problem at first sight, it actually makes the underlying physical mechanisms clearer. 
%
It turned out that even when the effective Richardson number (see (\ref{netJ})) is greater than 1/4 everywhere, in order to guarantee the stability of the flow the following two conditions must be satisfied; (i) the directions of the shear and the stratification are perfectly aligned, and (ii) there is no external magnetic field. If either of the conditions is not met, only an upper bound of the growth rate can be found.
The conclusion obtained for non-magnetised flows is consistent with the observation by Candelier et al. (2011), where the stability of the two-dimensional plane Bickley jet is studied. 
Linear and nonlinear dynamics of flows having a horizontal component of shear under vertical gravity has attracted much attention in hydrodynamic stability studies in recent years (e.g. Chen et al. (2016), Lucas et al. (2017), Facchini et al. (2018)). 
It is an interesting question what the upper bound of the growth rate found in this paper imply for those problems.
Moreover, the discovery that the Miles-Howard type stabilisation mechanism might be not so robust in magnetised problems may have some implications for solar physics and engineering problems.
For example, the Richardson number in the tachocline is known to be much larger than a quarter (see Cally (2000)), but it is not surprising that ideal instabilities occur there owing to the much magnetic fields present.

The second bound (theorem 2) can be regarded as a generalised version of the semi-ellipse theorem by Kochar \& Jain (1979), including the compressible, magnetohydrodynamic and non-planar effects. The basic idea used in this bound is to use the integral derived in the generalised Miles-Howard theory in order to tighten the semicircle bound. 
A notable feature of the second bound is that, unlike the first bound, it can incorporate the stabilisation effect of stable stratification.
However, consistent with the caveat found for the Miles-Howard stability condition, the existence of the shear perpendicular to the direction of the stratification reduces the amount of the improvement on the bound. 
The existence of a strong applied magnetic field also hinders the improvement, in which case the first bound may give a better result.

The third bound (theorem 3) is a generalisation of the eigenvalue bound found by Cally (1983), but here a quite different approach is used in the derivation.
From the generalised Miles-Howard theory we can deduce an inequality that must be satisfied somewhere in the flow to generate some instability. 
At each point in the domain, the inequality defines a semicircle region in the complex plane. Since the eigenvalue must lie within one of those semicircles, the net bound can be found by drawing the outer envelope of them.
The basic feature of this bound is that it gives a reasonably tight growth rate bound when the wavenumber is not too small, and the base shear is not too strong.





The bounds are compared with the numerical eigenvalues of 
the two canonical planar model flows used in the solar physics community (Bogdan \& Cally 1997; Tobias \& Hughes 2004). 
The first model flow concerns the Kelvin-Helmholtz type instability in a convectively stable polytrope atmosphere subjected to a uniform magnetic field.
As the instability is predominantly generated by the shear, when the base velocity is sufficiently reduced the growth rate must be suppressed. 
The inner envelope bound is particularly useful to capture this typical behaviour of the eigenvalue, even leading to a reasonable threshold value of the base velocity below which the flow is stable.
In the second model, the magnetic field is non-uniform, and  
the magnetic buoyancy instability occurs in the flow. 
For this case, three-dimensional perturbations are more dangerous and interesting, but such perturbations have not been taken into account in the previous complex eigenvalue bound analyses (Adam (1978b), Cally (1983)).
Theorems 1 and 3 derived in this paper successfully give the bounds
which are consistent with the numerical eigenvalues. 
\textcolor{black}{
All the three bounds are dependent on the wavelength of the perturbation, and becomes not so effective when it is very long. However, this is not really a practical disadvantage because the wavelength does not exceed the size of the system, so we can think of a maximum wavelength.
%
%
For example in Theorem 1, if we exclude long-wavelength disturbances, the bound should be better than all the previously known results. 
}

%

As remarked in section 1, an ideal instability of the streak plays a key role in the self-sustaining mechanism of nonlinear coherent structures in turbulent flows (Hall \& Smith 1991; Waleffe 1997; Wang et al. 2007; Hall \& Sherwin 2010; Deguchi \& Hall 2014). 
The streak is essentially the streamwise non-planar mean field, and hence 
the self-sustaining process has some resemblance to the mean-field theory in magnetohydrodynamics. 
The corresponding self-sustainment processes of coherent structures in the nonlinear magneto-rotational instability, the shear-driven dynamo, and the subcritical magneto-shear instability have recently been formulated by Riols et al. (2013), Deguchi (2019ab), and Deguchi (2020), respectively. 
The upper bound on the growth rate obtained in this paper may be useful in estimating how fast the turbulent burst will occur in the magneto-hydrodynamic flows.
The fact that the stability of the mean field is involved in the sustainment process of coherent structures implies that the results in this study may have some relevance for turbulence control technologies. Dong et al. (2012) in fact reported that turbulence can be suppressed by controlling the instability of the mean-field via the imposed streamwise magnetic field. 
The stability condition by the inner envelope theorem may allow us to know when such suppression will be activated.

Finally, we remark that Rayleigh's inflection point theorem is more difficult to apply for the generalised non-planar flows, unlike Howard's semicircle theorem and the Miles-Howard stability theory studied in this paper.
This is because the inflection point theorem essentially comes from the conservation of \textcolor{black}{enstrophy} that cannot be used for three-dimensional flows (see Arnold 1978; Dowling 1995).  
It would be worth noting that recently Deguchi (2019c) nevertheless derived a simple condition for the existence of a neutral mode for flows sheared in two directions. This is an extension of the result by Lin (1955), who showed that the inflection point condition can also be used to check the existence of a neutral mode in the classical shear flow problem. 
The key simplification here is that for neutral perturbations we only need to analyse the property of the flow around the critical layer at which the wave becomes singular. For compressible, stratified magneto-shear flows the structure of the singular layer is expected to become much more complicated, and so it is of interest to see how the condition should be modified.

The author wishes to deeply acknowledge Professor P. Cally for illuminating discussions and his numerical eigenvalue results used in section 5. The useful comments made by the referees should also be thanked.
This work was supported by Australian Research Council Discovery Early Career Researcher Award DE170100171.

The author reports no conflict of interest.

\appendix
\section{The derivation of the pressure equation}


The governing equations (2.5) can be reduced to a single equation for $\widetilde{q}$ as follows.
Denoting the divergence of the displacement vector as $\mathcal{D}=i\alpha \xi+\eta_y+\zeta_z$, from (2.8c) and (2.9a) the link between $\eta,\zeta,\widetilde{q}$ and $\mathcal{D}$ can be found as
\begin{eqnarray}
(s^2+a^2)\Lambda_c\mathcal{D}=-U^2\widetilde{q}-\Lambda_a(G_y\eta+G_z\zeta),\label{DD}
\end{eqnarray}
where
\begin{eqnarray}
\Lambda_c\equiv  \overline{\rho}(U^2-c_T^2),\qquad \Lambda_a\equiv  \overline{\rho}(U^2-a^2).
\end{eqnarray}
Here $c_T$ is the local cusp (tube) wave speed defined in section 3.2.
Equation (\ref{DD}) can then be employed to 
eliminate $\mathcal{D}$ and $\widetilde{\rho}$ from (2.8b), (2.9b) and (2.9c):
\begin{eqnarray}
\left [
\begin{array}{cc}
k^2\Lambda_a-\overline{\rho}\mathcal{N}_{1}^2 
& -\overline{\rho}\mathcal{N}_{12}^2  \\
-\overline{\rho}\mathcal{N}_{12}^2 
& k^2\Lambda_a-\overline{\rho}\mathcal{N}_{2}^2 
\end{array}
\right ]
\left [
\begin{array}{c}
\eta \\
\zeta
\end{array}
\right ]
=
\left [
\begin{array}{c}
\widetilde{q}_y-\frac{G_y\overline{\rho}U^2\widetilde{q}}{(s^2+a^2)\Lambda_c}\\
\widetilde{q}_z-\frac{G_z\overline{\rho}U^2\widetilde{q}}{(s^2+a^2)\Lambda_c}
\end{array}
\right ].\label{qetazeta}
\end{eqnarray}
The components of the matrix are defined using
\begin{eqnarray*}
\mathcal{N}_{1}^2 \equiv N_1^2+\frac{\overline{\rho}U^2}{\Lambda_c}\frac{c_T^2G_y^2}{s^4},\qquad
\mathcal{N}_{2}^2 \equiv N_2^2+\frac{\overline{\rho}U^2}{\Lambda_c}\frac{c_T^2G_z^2}{s^4}, \qquad
\mathcal{N}_{12}^2 \equiv N_{12}^2+\frac{\overline{\rho}U^2}{\Lambda_c}\frac{c_T^2G_yG_z}{s^4}.
\end{eqnarray*}
Finally, we use 
\begin{eqnarray}
0=(U^2-s^2)\widetilde{q}+(s^2+a^2)\Lambda_c(\eta_y+\zeta_z)+U^2\overline{\rho}(G_y\eta+G_z\zeta) \label{xiremoved}
\end{eqnarray}
that can be found by eliminating $\xi$ from (2.8c) and (2.9a). Equation (\ref{xiremoved}) becomes a single equation for $\widetilde{q}$ when $\eta$ and $\zeta$ are expressed by $\widetilde{q}$ using (\ref{qetazeta}). 

\section{The Euler-Lagrange equations}

The Euler-Lagrange equations associated with the optimisation problem (3.14) are
\begin{subequations}\label{ELeq}
\begin{eqnarray}
R^2\xi=(\overline{u}-r_c)^2\xi-(\mathbf{l}_1^{\dagger}\mathbf{x})/\overline{\rho}k^{2},\\
R^2\eta=(\overline{u}-r_c)^2\eta-\{\mathbf{l}_2^{\dagger}\mathbf{x}-\partial_y (\mathbf{l}_4^{\dagger}\mathbf{x})\}/\overline{\rho}k^{2},\\
R^2\zeta=(\overline{u}-r_c)^2\zeta-\{\mathbf{l}_3^{\dagger}\mathbf{x}-\partial_z (\mathbf{l}_4^{\dagger}\mathbf{x})\}/\overline{\rho}k^{2}.
\end{eqnarray}
\end{subequations}
In principle, for fixed $r_c$, the optimised value $R^2$ can be found by solving those equations 
using some numerical eigenvalue solver.
Here $\mathbf{x}$ is the transpose of $[\xi,\eta,\zeta,\eta_y+\zeta_z]$ and
\begin{subequations}\label{LLLL}
\begin{eqnarray}
\mathbf{l}_1^{\dagger}=\overline{\rho} [k^2s^2,-ik G_y,-ik G_z,-ik s^2],\\
\mathbf{l}_2^{\dagger}=\overline{\rho} [ik G_y,k^2a^2+G_y\frac{\overline{\rho}_y}{\overline{\rho}},G_y\frac{\overline{\rho}_z}{\overline{\rho}}, G_y ], \\
\mathbf{l}_3^{\dagger}=\overline{\rho} [ik G_z,G_z\frac{\overline{\rho}_y}{\overline{\rho}}, k^2a^2+G_z\frac{\overline{\rho}_z}{\overline{\rho}}, G_z],\\
\mathbf{l}_4^{\dagger}=\overline{\rho} [ik s^2,G_y,G_z,s^2+a^2].
\end{eqnarray}
\end{subequations}
The daggers describe a Hermitian transpose. As commented in the main text, calculating $R^2$ in this way is inefficient.

The computational cost could be reduced by using the fact that the terms in the energy equation can be written in a quadratic form; there exists the Hermitian matrix $\mathbb{M}$ such that
\begin{eqnarray}
\langle (\overline{u}-r_c)^2 Q \rangle-\langle \mathcal{L} \rangle- R^2\langle Q\rangle=\langle \overline{\rho}\mathbf{x}^{\dagger}\mathbb{M}\mathbf{x}\rangle.\label{xMx}
\end{eqnarray}
A straightforward algebra yields
\begin{eqnarray}
\mathbb{M}=\overline{\rho}k^2\{(\overline{u}-r_c)^2-R^2\}\text{diag}(1,1,1,0)-
\left [
\begin{array}{c}
\mathbf{l}_1^{\dagger}\\
\mathbf{l}_2^{\dagger}\\
\mathbf{l}_3^{\dagger}\\
\mathbf{l}_4^{\dagger}
\end{array}
\right ],
\end{eqnarray}
using (\ref{LLLL}).
If the values of $R^2,r_c$ are given, the four eigenvalues of this matrix ($\lambda_i(y,z)$, $i=1,2,3,4$ say) can be easily found numerically at each point $(y,z)$. Then the minimum value of $R^2$ that realises
\begin{eqnarray}
0=\max_{\Omega}\{\max (\lambda_1,\lambda_2,\lambda_3,\lambda_4)\}
\end{eqnarray}
gives the radius of our interest.
The above condition ensures the negative definiteness of $\mathbf{x}^{\dagger}\mathbb{M}\mathbf{x}$, and therefore (3.13) follows using (\ref{xMx}). 
Note that the eigenvalue bound found in the matrix method may be looser than the Euler-Lagrange bound, because in the former method the link between $\eta, \zeta, $ and $ \eta_y+\zeta_z$ is lost. This matrix idea can be further advanced to yield the analytic bound summarised in Theorem 1.




\section{Optimisation of $\sigma$ in the inner envelope bound}

The best inner envelope bound can be obtained by choosing $\sigma\in (0, s^2+a^2]$ that minimises $\lambda \equiv (\lambda_2+k^{-1}\lambda_1+k^{-2}\lambda_0)$ at each point $y,z$.
At $\sigma=s^2$ the largest eigenvalue $\lambda_1$ changes its form (see (3.16)) so we must consider two intervals $0<\sigma < s^2$ and $s^2\leq \sigma \leq s^2+a^2$ separately; 
we shall shortly see that the minimum of $\lambda$ can always be obtained by the second interval.

Hereafter we denote
\begin{eqnarray}
\widehat{N}_1^2\equiv \frac{G_y\overline{\rho}_y}{\overline{\rho}}-\frac{G_y^2}{\sigma},~~~
\widehat{N}_2^2\equiv \frac{G_z\overline{\rho}_z}{\overline{\rho}}-\frac{G_z^2}{\sigma}.
\end{eqnarray}
For the interval $s^2\leq \sigma \leq s^2+a^2$, the largest eigenvalues are found as
\textcolor{black}{
\begin{eqnarray}
\lambda_2=(\overline{u}-r_c)^2-s^2(1-\frac{s^2}{\sigma}),\\
\lambda_1=(1-\frac{s^2}{\sigma})\sqrt{G_y^2+G_z^2},\\
\lambda_0=\max(0,-(\widehat{N}_1^2+\widehat{N}_2^2)).
\end{eqnarray}
}
Thus 
\begin{eqnarray}
\lambda(\sigma)=
\left \{
\begin{array}{c}
\lambda_+(\sigma)\qquad \text{if} \qquad \widehat{N}_1^2+\widehat{N}_2^2\geq 0,\\
\lambda_-(\sigma)\qquad \text{if} \qquad \widehat{N}_1^2+\widehat{N}_2^2<0,
\end{array}
\right .\label{lambdasigma}
\end{eqnarray}
where
\textcolor{black}{
\begin{subequations}\label{lambda12}
\begin{eqnarray}
\lambda_+(\sigma)\equiv(\overline{u}-r_c)^2+s^2\left (1-\frac{s^2}{\sigma}\right )\left (\frac{k_h}{k}-1 \right ),\hspace{20mm}\\
\label{lambda1}
\lambda_-(\sigma)\equiv (\overline{u}-r_c)^2
+s^2\left (1-\frac{s^2}{\sigma}\right )\left (\frac{k_h}{k}-1 \right )+\frac{s^4}{\sigma}\frac{k^2_h}{k^2}-\frac{G_y\overline{\rho}_y+G_z\overline{\rho}_z}{\overline{\rho}k^2}.
\label{lambda2}
\end{eqnarray}
\end{subequations}
}
Here $k_h$ is the wavenumber associated with the density scale height defined in section 3.1.
From (\ref{lambda12}) we can show for any $\sigma_1<\sigma_2$ that 
\begin{subequations}
\begin{eqnarray}
\lambda_+(\sigma_1)<\lambda_+(\sigma_2),~~ \lambda_-(\sigma_1)>\lambda_-(\sigma_2),\qquad \text{if} ~~k<k_h,\label{kkh1}\\
\lambda_+(\sigma_1)>\lambda_+(\sigma_2),~~ \lambda_-(\sigma_1)>\lambda_-(\sigma_2),\qquad \text{if} ~~k>k_h.\label{kkh2}
\end{eqnarray}
\end{subequations}
For the range of $\sigma$ under consideration, $N_1^2 \leq \widehat{N}_1^2 \leq N_{1a}^2$ and $N_2^2 \leq \widehat{N}_2^2 \leq N_{2a}^2$, where $N_{1a}^2, N_{2a}^2$ are buoyancy frequencies defined in section 3.1.

\begin{table}
  \begin{center}
    \begin{tabular}{ccc}
  ~  & $k<k_h$ & $k\geq k_h$  \\ 
   $0<N_1^2+N_2^2<N_{1a}^2+N_{2a}^2$ & 0 &  $c_T^2(\frac{k_h}{k}-1)$ \\
    $N_1^2+N_2^2<0<N_{1a}^2+N_{2a}^2$ & $-\frac{N_1^2+N_2^2}{k_h^2}(\frac{k_h}{k}-1)$ & $c_T^2(\frac{k_h}{k}-1)$  \\ 
    $N_1^2+N_2^2<N_{1a}^2+N_{2a}^2<0$ & $c_T^2(\frac{k_h}{k}-1)-\frac{N_{a1}^2+N_{a2}^2}{k^2}$ & $c_T^2(\frac{k_h}{k}-1)-\frac{N_{a1}^2+N_{a2}^2}{k^2}$ 
    \end{tabular}
  \end{center}
\caption{The summary of the optimised values of $\lambda-(\overline{u}-r_c)^2$. } 
\label{sample-table}
\end{table}

When $k>k_h$, (\ref{kkh2}) implies that the larger the value of $\sigma$, the smaller the associated value of $\lambda$. This means that the best choice of $\sigma$ is the largest possible value $s^2+a^2$, without regarding the sign of $(\widehat{N}_1^2+\widehat{N}_2^2)$. Thus, noting $\lambda_+(s^2+a^2)=(\overline{u}-r_c)^2+c_T^2\left (\frac{k_h}{k}-1 \right )$ and $\lambda_-(s^2+a^2)=(\overline{u}-r_c)^2+c_T^2\left (\frac{k_h}{k}-1 \right )-k^{-2}(N_{1a}^2+N_{2a}^2)$, we have the optimums summarised at the rightmost column of Table 1. 

When $k<k_h$, the situation is more complicated as expected from (\ref{kkh1}).
Let $s^2\leq \sigma \leq s^2+a^2$ and $k<k_h$. We shall deduce $\sigma$ that gives the minimum value of $\lambda$.
Here it is convenient to introduce $\widehat{N}^2(\sigma)=\widehat{N}_1^2+\widehat{N}_2^2$. Of course, $\widehat{N}^2(s^2)=N^2$ and $\widehat{N}^2(s^2+a^2)=N_{a}^2$, writing $N^2=N_1^2+N_2^2$ and $N_{a}^2=N_{1a}^2+N_{2a}^2$. Moreover, $\widehat{N}^2(\sigma)=0$ when $\sigma=\sigma_0\equiv \frac{\overline{\rho}s^4k_h^2}{G_y\overline{\rho}_y+G_z\overline{\rho}_z}$.
Depending on the sign of $N^2$ and $N_{a}^2$, following three cases are possible.
\begin{enumerate}
\item When $0<N^2<N_{a}^2$, $\widehat{N}^2(\sigma)=\widehat{N}_1^2+\widehat{N}_2^2$ is positive. The optimum must be found by $\lambda_+$, and thus from (\ref{kkh1}) we select the smallest possible value $\sigma=s^2$. Noting $\lambda_+(s^2)=(\overline{u}-r_c)^2$, we have the result shown in Table 1.
\item When $N^2<N_{a}^2<0$, $\widehat{N}^2(\sigma)=\widehat{N}_1^2+\widehat{N}_2^2$ is negative. The optimum must be found by $\lambda_-$, and thus from (\ref{kkh1}) we select the largest possible value $\sigma=s^2+a^2$. The result is unchanged from the $k\geq k_h$ case, as shown in Table 1.
\item When $N^2<0<N_{a}^2$, we need to split the interval of $\sigma$ into two parts using $\sigma_0$. For $s^2<\sigma<\sigma_0$, $\widehat{N}(\sigma)<0$ and thus $\lambda_-$ must be used to compute the optimum, while for $\sigma_0<\sigma<s^2+a^2$, $\widehat{N}(\sigma)>0$ and we should use $\lambda_+$. From (\ref{kkh1}) the optimum is $\lambda_-(\sigma_0)=\lambda_+(\sigma_0)=-\frac{N^2}{k_h^2}\left (\frac{k_h}{k}-1 \right )$. Note that $c_T^2-\frac{N_{a}^2}{k_h^2}=-\frac{N^2}{k_h^2}$ and thus the optimum changes continuously; see Table 1.
\end{enumerate}


Next we show that the consideration of the other interval $0< \sigma \leq s^2$ does not change the optimum values given in Table 1. From (\ref{eigen12}) the explicit expression of $\lambda$ becomes (\ref{lambdasigma}) with
\begin{subequations}\label{lambda12b}
\begin{eqnarray}
\lambda_+(\sigma)\equiv(\overline{u}-r_c)^2-s^2\left (1-\frac{s^2}{\sigma}\right )\left (\frac{k_h}{k}+1 \right ),\hspace{20mm}\\
\label{lambda1b}
\lambda_-(\sigma)\equiv (\overline{u}-r_c)^2
-s^2\left (1-\frac{s^2}{\sigma}\right )\left (\frac{k_h}{k}+1 \right )+\frac{s^4}{\sigma}\frac{k^2_h}{k^2}-\frac{G_y\overline{\rho}_y+G_z\overline{\rho}_z}{\overline{\rho}k^2}.
\label{lambda2b}
\end{eqnarray}
\end{subequations}
Clearly, the optimum can always be found by the largest possible value $\sigma=s^2$. 
This means that the optimised value is $\lambda-(\overline{u}-r_c)^2=-\frac{1}{k^2}\min(N^2,0)$. It is easy to see this optimum value is larger than those found in Table 1, noting the identities
\begin{eqnarray}
-\frac{N^2}{k_h^2}(\frac{k_h}{k}-1)+\frac{N^2}{k^2}=-\frac{N^2}{k_h^2}(\frac{k_h}{k}-1-\frac{k_h^2}{k^2})\\
c_T^2(\frac{k_h}{k}-1)-\frac{N_a^2}{k^2}+\frac{N^2}{k^2}
=\frac{N_a^2-N^2}{k_h^2}(\frac{k_h}{k}-1-\frac{k_h^2}{k^2}),
\end{eqnarray}
and the inequality $(\frac{k_h}{k}-1-\frac{k_h^2}{k^2})<0$.

Using Table 1 noting that $-\frac{N_1^2+N_2^2}{k_h^2}=c_T^2-\frac{N_{1a}^2+N_{2a}^2}{k_h^2}$, we arrive at Theorem 1.

\section{Derivation of the integral in section 4.1}

Integrating $\phi^*\times$(4.1c)+$\psi^*\times$(4.1d) by parts over the domain
noting 
$(U^{-1/2}\phi^*)_y\widetilde{q}+(U^{-1/2}\psi^*)_z\widetilde{q}=U^{-1/2}(\widehat{\kappa}_1^*-U^{-1}\widehat{\kappa}_2^*)\widetilde{q}$,
and eliminating $\widetilde{q}$ and $\varphi$ using
\begin{eqnarray}
 \varphi
=
\frac{ (s^2\widehat{\kappa}
+\widehat{\mathcal{G}})}{ik (U^2-s^2)},\qquad
ik \varphi +\widehat{\kappa}=\frac{U^2\widehat{\kappa}+\widehat{\mathcal{G}}}{U^2-s^2},
\\
U^{1/2}\widetilde{q}=
-\frac{ \overline{\rho}(a^2+s^2)(U^2-c_T^2)\widehat{\kappa}}{U^2-s^2}
-\frac{ U^2\overline{\rho}\widehat{\mathcal{G}}}{U^2-s^2},
\end{eqnarray}
derived by (4.1a) and (4.1b), we arrive at 
\begin{eqnarray}
0= \left \langle k^2\overline{\rho}\frac{U^2-a^2}{U}(|\phi|^2+|\psi|^2)-\frac{G_y}{U\overline{\rho}_y}|\overline{\rho}_y\phi+\overline{\rho}_z\psi|^2 \right . \nonumber \\
-\overline{\rho}\frac{a^2}{U}\left (|\widehat{\kappa}_1|^2+\frac{|\widehat{\kappa}_2|^2}{U^2}-\frac{\widehat{\kappa}_2^*\widehat{\kappa}_1+\widehat{\kappa}_2\widehat{\kappa}_1^*}{U} \right )\nonumber \\
-\overline{\rho}\frac{s^2U}{U^2-s^2}\left(|\widehat{\kappa}_1|^2+\frac{|\widehat{\kappa}_2|^2}{U^2}-\frac{\widehat{\kappa}_2^*\widehat{\kappa}_1+\widehat{\kappa}_2\widehat{\kappa}_1^*}{U}\right )\nonumber \\
\left. -\frac{U\overline{\rho}\{(\widehat{\kappa}_1^*-U^{-1}\widehat{\kappa}_2^*)\widehat{\mathcal{G}}+\widehat{\mathcal{G}}^*(\widehat{\kappa}_1-U^{-1}\widehat{\kappa}_2)\}}{U^2-s^2}-\frac{\overline{\rho}|\widehat{\mathcal{G}}|^2}{U(U^2-s^2)}
 \right \rangle.\label{appendeq}
\end{eqnarray}
Let us extract the imaginary part of the integrand. The imaginary parts of the terms in the first to forth lines above can be found as 
\begin{subequations}
\begin{eqnarray}
-k^2\overline{\rho}(1+\frac{a^2}{|U|^2})(|\phi|^2+|\psi|^2)-\frac{G_y}{|U|^2\overline{\rho}_y}|\overline{\rho}_y\phi+\overline{\rho}_z\psi|^2,~~~\label{line1}\\
-\overline{\rho}a^2\frac{|\widehat{\kappa}_1|^2}{|U|^2}-\overline{\rho}a^2\frac{(4U_r^2-|U|^2)|\widehat{\kappa}_2|^2}{|U|^6}+\overline{\rho}a^2\frac{2U_r(\widehat{\kappa}_2^*\widehat{\kappa}_1+\widehat{\kappa}_2\widehat{\kappa}_1^*)}{|U|^4},~~~\label{line2}\\
-\overline{\rho}s^2\frac{(|U|^2+s^2)|\widehat{\kappa}_1|^2}{|U^2-s^2|^2}
+\overline{\rho}s^2\frac{2U_r(\widehat{\kappa}_2^*\widehat{\kappa}_1+\widehat{\kappa}_2\widehat{\kappa}_1^*)}{|U^2-s^2|^2}
+\overline{\rho}s^2\frac{(|U|^2+s^2-4U_r^2)|\widehat{\kappa}_2|^2}{|U|^2|U^2-s^2|^2},~~~\label{line3}\\
-\overline{\rho}\frac{(|U|^2+s^2)(\widehat{\kappa}_1^*\widehat{\mathcal{G}}+\widehat{\mathcal{G}}^*\widehat{\kappa}_1)}{|U^2-s^2|^2}
+\overline{\rho}\frac{2U_r(\widehat{\kappa}_2^*\widehat{\mathcal{G}}+\widehat{\mathcal{G}}^*\widehat{\kappa}_2)}{|U^2-s^2|^2}
+\overline{\rho}\frac{(|U|^2+s^2-4U_r^2)|\widehat{\mathcal{G}}|^2}{|U|^2|U^2-s^2|^2},~~~\label{line4}
\end{eqnarray}
\end{subequations}
respectively. Here $U_r$ represents the real part of $U$.

The terms in (\ref{line2}) can be transformed into 
\begin{eqnarray}
-\overline{\rho}\frac{a^2}{|U|^2}|\widehat{\kappa}-\frac{2U_r}{|U|^2}\widehat{\kappa}_2|^2+\overline{\rho}\frac{a^2}{|U|^4}|\widehat{\kappa}_2|^2,
\end{eqnarray}
while the summation of the terms shown in (\ref{line3}) and (\ref{line4}) becomes
\begin{eqnarray}
-\overline{\rho}s^2\frac{|U|^2+s^2}{|U^2-s^2|^2}\left |\widehat{\kappa}_1-\frac{2U_r}{|U|^2+s^2}\widehat{\kappa}_2+\frac{\widehat{g}}{s^2}\right |^2 \nonumber \\
+\overline{\rho}\frac{s^2|\widehat{\kappa}_2|^2}{|U^2-s^2|^2}\left (\frac{4U_r^2}{|U|^2+s^2}+\frac{|U|^2+s^2-4U_r^2}{|U|^2|U^2-s^2|^2}\right ) \nonumber \\
+\overline{\rho}\frac{|\widehat{\mathcal{G}}|^2}{|U^2-s^2|^2}\left (\frac{|U|^2+s^2}{s^2}+\frac{|U|^2+s^2-4U_r^2}{|U|^2}\right ).\label{A2cd}
\end{eqnarray}
Further applying the identities
\begin{subequations}
\begin{eqnarray}
\frac{4U_r^2}{|U|^2+s^2}+\frac{|U|^2+s^2-4U_r^2}{|U|^2}
=
\frac{|U^2-s^2|^2}{|U|^2(|U|^2+s^2)},\\
\frac{|U|^2+s^2}{s^2}
+\frac{|U|^2+s^2-4U_r^2}{|U|^2}
=\frac{|U^2-s^2|^2}{|U|^2s^2},
\end{eqnarray}
\end{subequations}
to (\ref{A2cd}), equation (4.3) in the main text follows. 


\section{Lower bound estimation of the buoyancy related term in section 4.2}
In order to use the result in section 4.1, we need to link $Q_1$ and $Q_2$. 
Writing $\widehat{\kappa}_1=U^{1/2}\kappa+U^{-1/2}\kappa_2$ with $\kappa_2=U^{-1/2}\widehat{\kappa}_2=(U_y\eta+U_z\zeta)/2$, we obtain the identity
\begin{eqnarray}
\widehat{\kappa}_1-\frac{2U_r}{|U|^2+s^2}\widehat{\kappa}_2+\frac{\widehat{\mathcal{G}}}{s^2}=
U^{1/2}(\kappa+\frac{\mathcal{G}}{s^2})+U^{-1/2}\kappa_2(1-\frac{2U_rU}{|U|^2+s^2}).
\end{eqnarray}
From the elemental inequality $f_1f_2^*+f_1^*f_2\leq 2|f_1||f_2|$ for any complex values $f_1,f_2$, we can find the following estimates for the terms appeared in (4.4):
\begin{eqnarray}
\left |\widehat{\kappa}_1-\frac{2U_r}{|U|^2+s^2}\widehat{\kappa}_2+\frac{\widehat{\mathcal{G}}}{s^2} \right |^2~~~~~~~~~~~~~~~~~~~~~~~~~~~~~~~~~~~~~~~~~~~~~~~~~~~~~~~~~~\nonumber \label{ineqQ11}\\
\geq
 |U| \left |\kappa+\frac{\mathcal{G}}{s^2} \right |^2+\frac{|U^2-s^2|^2}{(|U|^2+s^2)^2}\frac{| \kappa_2|^2}{|U|}-2\frac{|U^2-s^2|}{|U|^2+s^2}| \kappa_2| \left |\kappa+\frac{\mathcal{G}}{s^2} \right |,\\
\left |\widehat{\kappa}_1-\frac{2U_r}{|U|^2}\widehat{\kappa}_2 \right |^2\geq
 |U||\kappa|^2+\frac{| \kappa_2|^2}{|U|}-2| \kappa_2| |\kappa|.~~~~~~~~~~~~~~~~~~~~~~~~~~~~~~~~~~\label{ineqQ12}
\end{eqnarray}
Here the identity $|U^2-s^2|^2=(|U|^2+s^2)^2-4U_r^2s^2$ may be useful to derive the first inequality.
The inequalities (\ref{ineqQ11}) and (\ref{ineqQ12}) can be used to show that
\begin{eqnarray*}
Q_1\geq |U|Q_2+\frac{\overline{\rho}|\widehat{\kappa}_2|^2}{|U|}\left (\frac{s^2}{|U|^2+s^2}+\frac{a^2}{|U|^2} \right )-2\overline{\rho}|\kappa_2|\frac{s^2|\kappa+s^{-2}\mathcal{G}|}{|U^2-s^2|}-2\overline{\rho}\frac{a^2}{|U|^2}|\kappa_2||\kappa|.\label{estQ1}
\end{eqnarray*}

The above estimate can be combined with (4.3) to yield
\begin{eqnarray}
0\geq \left \langle |U|Q_2+\frac{f_b}{|U|}-\frac{\overline{\rho}}{|U|}\frac{a_M^2}{c_i^2}|\kappa_2|^2-2\overline{\rho}|\kappa_2|\frac{s^2|\kappa+s^{-2}\mathcal{G}|}{|U^2-s^2|} \right \rangle. \label{coQ2in}
\end{eqnarray}
Here it is convenient to write
\begin{eqnarray}
\mathcal{A}\equiv \sqrt{\langle |U|Q_2 \rangle},\qquad
\mathcal{B}\equiv \sqrt{\left \langle \frac{4\overline{\rho}}{|U|}|\kappa_2|^2 \right \rangle},\qquad
\mathcal{F}\equiv \frac{1}{\mathcal{B}^2}\left \langle \frac{f_b}{|U|}-\frac{\overline{\rho}}{|U|}\frac{a_M^2}{c_i^2}|\kappa_2|^2 \right \rangle.\label{defAB}
\end{eqnarray} 
Noting that $\langle 2\overline{\rho}|\kappa_2|\frac{s^2|\kappa+s^{-2}g|}{|U^2-s^2|}\rangle \leq \mathcal{A}\mathcal{B}$ holds because of the Schwarz inequality, the inequality (\ref{coQ2in}) can be written in the simple form 
\begin{eqnarray}
0\geq \mathcal{A}^2-\mathcal{A}\mathcal{B}+\mathcal{B}^2\mathcal{F}.\label{AAABF}
\end{eqnarray} 

Next we shall see how the above inequality can be used to estimate the buoyancy term.
Let us consider
\begin{eqnarray}
\frac{\widehat{f}_b}{\overline{\rho}}-4J_m|\widehat{\kappa}_2|^2=[\phi^*, \psi^*]
\left [
\begin{array}{cc}
\overline{u}_y^2(J_1-J_m) & \overline{u}_y\overline{u}_z(J_{12}-J_m) \\
\overline{u}_y\overline{u}_z(J_{12}-J_m) & \overline{u}_z^2(J_2-J_m)
\end{array}
\right ]
\left [
\begin{array}{c}
\phi\\ \psi
\end{array}
\right ]
,~~~\label{phipsiq2}
\end{eqnarray}
which is similar to (\ref{phipsiq1}). 
The two eigenvalues $\lambda_+,\lambda_-$ of the matrix in (\ref{phipsiq2}) are found as
\begin{eqnarray}
\lambda_{\pm}=\frac{(\overline{u}_y^2+\overline{u}_z^2)(J-J_m)\pm \sqrt{(\overline{u}_y^2+\overline{u}_z^2)^2(J-J_m)^2-4\overline{u}_y^2\overline{u}_z^2J_m(2J_{12}-J_1-J_2)}}{2}.
\end{eqnarray}
Therefore, 
\begin{eqnarray}
\widehat{f}_b-4J_m\overline{\rho}|\widehat{\kappa}_2|^2+\lambda_M\overline{\rho}(|\phi|^2+|\psi|^2)\geq 0,\label{alwayspos}
\end{eqnarray} 
where $\lambda_M\equiv\max_{\Omega}|\lambda_-|$ is the quantity defined in (4.27).
Integrating this inequality over $\Omega$ and using (4.22), we can deduce
\begin{eqnarray}
\langle f_b \rangle  \geq \langle 4J_m\overline{\rho}|\kappa_2|^2-\lambda_M\overline{\rho}(|\eta|^2+|\zeta|^2)\rangle \geq 4J_m\langle \overline{\rho}|\kappa_2|^2\rangle-\frac{\lambda_M}{k^2}\langle Q_2\rangle. \label{fbQ2bound}
\end{eqnarray}
This is essentially the estimate of the buoyancy term (4.26) but we still need to find the relation between $\langle \overline{\rho}|\kappa_2|^2\rangle$ and $\langle Q_2\rangle$. For this purpose we use (\ref{AAABF}).

We note that by the definition of $\mathcal{A}$, $\mathcal{B}$ and $\mathcal{F}$ given in (\ref{defAB}), 
\begin{eqnarray}
\left (\mathcal{F}+\frac{a_M^2}{4c_i^2} \right )\mathcal{B}^2=\left \langle \frac{f_b}{|U|} \right \rangle \geq 4J_m \left \langle \frac{\overline{\rho}|\kappa_2|^2}{|U|} \right \rangle-\frac{\lambda_M}{k^2}\left \langle \frac{Q_2}{|U|} \right \rangle \geq J_m\mathcal{B}^2-\frac{\lambda_M}{k^2c_i^2}\mathcal{A}^2.
\end{eqnarray}
Here to estimate $f_b$ we have used (\ref{alwayspos}). 
Together with (\ref{AAABF}), the latter inequality becomes
\begin{eqnarray}
0\geq \mu \left (\frac{\mathcal{A}}{\mathcal{B}}\right )^2-\frac{\mathcal{A}}{\mathcal{B}}+J_m-\frac{a_M^2}{4c_i^2},\label{ABquad}
\end{eqnarray}
where $\mu$ is defined in (4.29).

Hereafter we consider unstable eigenvalues that make $\mu$ positive. 
In this case (\ref{ABquad}) implies that the quantity $\mathcal{A}/\mathcal{B}$ can be bounded from above.
In fact, since (\ref{ABquad}) can be rewritten as
\begin{eqnarray}
0\geq \mu \left (\frac{\mathcal{A}}{\mathcal{B}}-\frac{1}{2\mu} \right )^2-\frac{1}{4\mu}+J_m-\frac{a_M^2}{4c_i^2},\label{ABquad2}
\end{eqnarray}
we have the estimate
\begin{eqnarray}
\frac{\mathcal{A}}{\mathcal{B}} \leq \frac{1+\sqrt{1+\mu(\frac{a_M^2}{c_i^2}-4J_m)}}{2\mu}.\label{ABC2}
\end{eqnarray}
The right side of this inequality is $\mathcal{C}$ defined in (\ref{defCCC}).
Furthermore, the definitions of $\mathcal{A}$ and $\mathcal{B}$ (see (\ref{defAB})) imply
\begin{eqnarray}
\frac{\mathcal{A}^2}{\mathcal{B}^2}\geq \frac{c_i^2\langle Q_2 \rangle}{4\langle \overline{\rho}|\kappa_2|^2 \rangle}\label{ABQ2}
\end{eqnarray}
and thus we have the estimation of the buoyancy term (4.26)
from (\ref{ABC2}), (\ref{ABQ2}) and (\ref{fbQ2bound}). 


\section{Upper bound of $c_i^2$ for $J_m>\frac{1}{4}$ in Theorem 2}

By assumption, $J_m>0$. 
In view of (\ref{ABquad2}), if $\mu>0$ and $J_m> \frac{1}{4\mu}$ are satisfied
\begin{eqnarray}
c_i^2< \frac{a_M^2}{4J_m-\mu^{-1}}.
\label{ciJm}
\end{eqnarray}
The conditions for this inequality to be valid can be transformed as follows.
\begin{eqnarray}
J_m> \frac{1}{4\mu} ~~\text{and} ~~\mu>0 \iff \mu>\frac{1}{4J_m} \iff c_i^2(1-\frac{1}{4J_m})>k^{-2}\lambda_M.%
\label{ciJm2}
\end{eqnarray}
Note that the rightmost condition cannot be satisfied when $1-\frac{1}{4J_m}<0$ because $\lambda_M\geq 0$.
Thus the condition (\ref{ciJm2}) is equivalent to
\begin{eqnarray}
1-\frac{1}{4J_m}>0~~\text{and}~~c_i^2>\frac{k^{-2}\lambda_M}{1-\frac{1}{4J_m}}\equiv H_0.\label{ciJm3}
\end{eqnarray}

If (\ref{ciJm3}) is satisfied, we can use the inequality (\ref{ciJm}) which becomes
\begin{eqnarray}
(4J_m-1)c_i^4-(4J_mk^{-2}\lambda_M+a_M^2)c_i^2+k^{-2}\lambda_M a_M^2 <0,
\end{eqnarray}
from which we can deduce 
\begin{eqnarray}
H_-<c_i^2<H_+\label{HcH}
\end{eqnarray}
with
\begin{eqnarray}
H_{\pm}=\frac{(4J_m k^{-2}\lambda_M+a_M^2)\pm \sqrt{(4J_mk^{-2}\lambda_M+a_M^2)^2-4(4J_m-1)k^{-2}\lambda_M a_M^2}}{2(4J_m-1)}.
\end{eqnarray}
Note that
\begin{eqnarray}
H_{\pm}-H_0=\frac{\pm \sqrt{(4J_mk^{-2}\lambda_M-a_M^2)^2+4k^{-2}\lambda_M a_M^2}-(4J_mk^{-2}\lambda_M-a_M^2)}{2(4J_m-1)}
\end{eqnarray}
and thus $H_-<H_0<H_+$. 

Let $J_m>\frac{1}{4}$. Then $H_+$ gives the upper bound of $c_i^2$. We can prove this by contradiction. 
Suppose there exists an unstable eigenvalue satisfying $c_i^2\geq H_+$. 
Then, since $H_0<c_i^2$ the condition (\ref{ciJm3}) is met, and thus we must have (\ref{HcH}), which contradicts with the assumption.

\end{document}